\documentclass[useAMS,usenatbib,12pt,preprint]{mn2e}
\usepackage{graphicx,amssymb,wasysym}
\usepackage{natbib}
\usepackage{color}
\pdfoutput=1
\graphicspath{{figures/}}

\usepackage{hyperref}
\providecommand{\eprint}[1]{preprint (arXiv:\href{http://arxiv.org/abs/#1}{#1})}
\providecommand{\adsurl}[1]{\href{#1}{ADS}}

\newcommand{\msun}{M_{\astrosun}}
\def\r{\textit{\textbf{r}}}
\def\v{\textit{\textbf{v}}}
\def\k{\textit{\textbf{k}}}
\def\n{\textit{\textbf{n}}}

\title[II. Radioactively powered transients]
{The long-term evolution of neutron star merger remnants -- II. Radioactively powered transients}
\author[Grossman, et al.]{ Doron Grossman$^{1}$, Oleg Korobkin$^2$, Stephan Rosswog$^2$ and Tsvi Piran$^1$ 
\\
$^{1}$Racah Institute of Physics, The Hebrew University, Jerusalem 91904, Israel\\
$^2$Astronomy and Oskar Klein Centre, Stockholm University, AlbaNova, SE-10691 Stockholm, Sweden}

\begin{document}

\maketitle
\begin{abstract}
We use 3D hydrodynamic simulations of the long-term evolution
of neutron star merger ejecta to predict the light curves of electromagnetic transients
that are powered by the decay of freshly produced r-process nuclei. 
For the dynamic ejecta that are launched by tidal and hydrodynamic
interaction, we adopt grey opacities of 10~cm$^2$/g, as suggested by recent studies. 
For our reference case of a 1.3 -- 1.4~$\msun$ merger, we find a broad IR peak 2-4~d after the merger. 
The peak luminosity is $\approx 2\times 10^{40}$~erg/s for an average orientation, but 
increased by up to a factor of 4 for more favourable binary parameters and viewing angles. 
These signals are rather weak and hardly detectable within the large error box ($\sim 100$ deg$^2$) 
of a gravitational wave trigger. A second electromagnetic transient results from 
neutrino-driven winds. These winds  produce `weak' r-process material with
$50 < A < 130$ and  abundance patterns that vary substantially between different
merger cases. 
For an adopted opacity of 1~cm$^2$/g, the resulting transients peak in the UV/optical 
about 6~h after the merger with a luminosity of  $\approx 10^{41}$~erg/s (for a wind of 0.01~$\msun$)
These signals are marginally detectable in deep follow-up searches (e.g. using Hypersuprime camera on Subaru).
A subsequent detection of the weaker but longer lasting
IR signal would allow an identification of the merger event. We briefly discuss the implications of
our results to the recent detection of an nIR transient accompanying GRB~130603B.
\end{abstract}


\section{Introduction}
\label{sec:introduction}

The merger of two neutron stars (ns$^2$), or, alternatively, of an ns
with a stellar-mass black hole (nsbh), plays a key role for several astrophysical 
questions. Such mergers are in the main focus of current efforts 
towards direct gravitational wave (GW) detections, they are thought to be the `engines' 
of short gamma-ray bursts \citep[sGRBs; see][for reviews] {piran04,lee07,nakar07} and they 
may produce the heaviest elements in the Universe \citep{lattimer74,lattimer77,eichler89,freiburghaus99b}.
Until recently, each of these topics has been mostly studied in its own right,
and currently much effort is invested in integrating the different facets of
the topic into a coherent multi-messenger picture 
\citep*{bloom09b,phinney09,metzger12a,bartos13,kasliwal13,kelley12,nissanke13,rosswog13a,piran13a}

By the end of this decade the second generation of ground-based
GW detectors is expected to be running at their
intended design sensitivities \citep{harry10,accadia11,somiya12}.
The main target of these detectors is the discovery of the chirping 
GW signals from ns$^2$ or nsbh mergers. The 
expected detection horizons are a few hundred Mpc for ns$^2$ mergers 
and about a Gpc for nsbh mergers (445/927 Mpc are adopted by the LIGO-Virgo
collaboration as canonical values; \citealt{abadie10}).

It had been realized early on that an electromagnetic (EM) transient
coincident with the GW signal could deliver crucial complementary information
\citep{kochanek93,hughes03,dalal06,arun09}. A spatial and temporal 
coincidence between a GW and an expected EM signal would enhance the
confidence in a GW detection and thereby effectively enhance the 
sensitivity of existing GW detector facilities. GW detections can provide 
the parameters of the merging system such as the individual masses or the
orbital inclination, but with sky localizations of tens of degrees they
leave us essentially blind with respect to the astronomical environment of 
a merger. Additional EM transients can complement the picture by,
for example, delivering accurate source localizations or redshifts and they
can therefore provide independent pieces of evidence for the nature of 
the observed event. Moreover, even before the detector facilities are 
fully upgraded, EM transients could improve the poorly constrained merger rates.

sGRBs have been long recognized \citep{eichler89} 
as possible EM counterparts of compact binary mergers. Circumstantial 
evidence for this association arises from various lines of argument. For example, 
the overall sGRB rate \citep{guetta06,nakar06,guetta09,coward12} is comparable to the 
estimates for binary neutron stars \citep{narayan91,phinney91,kalogera04a,kalogera04b,abadie10} 
and the redshift distribution of sGRBs is consistent with sources that are delayed 
with respect to the star formation rate \citep{guetta06,nakar06,nakar07}. 
Further evidence comes from the location of some sGRBs within elliptical 
galaxies which harbor very little star formation and from the location of 
sGRBs with respect to their host galaxies \citep{berger09,berger10,fong10,fong13}. However, 
sGRBs are most likely beamed, both theoretical and observational studies point to
opening angles roughly around  5$^\circ$ \citep{rosswog03b,aloy05,fong12}, but with 
large uncertainties.
Most likely the beaming factor is large and in most cases the sGRB will be undetectable
since it points away from us.  In fact, since {\it Swift} has begun operating in 2004, no 
sGRB has been detected with a redshift below $z\approx 0.12$ \citep{rowlinson10}, 
while the envisaged horizons of the advanced detector facilities are $\approx 0.05/0.1$ 
for ns$^2$ and nsbh mergers.  

`Macronovae'\footnote{This term has been introduced to the literature by \cite{kulkarni05}, 
other authors prefer `kilonova' \citep{metzger10b}.} are more isotropic transients that are
powered by freshly synthesized radioactive isotopes and therefore are closely related 
to cosmic nucleosynthesis. 
After it had been realized that ns mergers eject $\sim 0.01 \msun$ 
\citep{rosswog98a,rosswog99} and that the ejecta consist essentially entirely 
of heavy ($A>130$) r-process matter \citep{freiburghaus99b},
\cite{li98} addressed the question how an ns merger would
look electromagnetically. In their first approach, they modelled the ejecta as 
a uniformly expanding sphere with a fixed opacity of $\kappa=0.2$ cm$^2$/g
and assumed that a fraction $f$ of the rest mass would be channelled into 
radioactive heating. Subsequently much effort has been invested in
refining macronova models and in further understanding their properties
\citep{kulkarni05,rosswog05a,metzger10b,goriely11a,roberts11,bauswein13a,piran13a,rosswog13a}.

All of these studies share the use of opacities that are characteristic for the line expansion
opacities of iron group elements. \cite{kasen13a}, however, recently pointed out that the opacities
of compact binary merger ejecta are likely dominated by elements with half-filled f-shells 
such as lanthanides and actinides which have extremely large opacities even if their mass 
fraction is as low as 1~per cent. Based on detailed atomic structure calculations they concluded 
that average opacities may be around 10~cm$^2$/g (rather than the previously used much 
lower values) and this would lead to weaker, later and redder emission \citep{barnes13a}. 
\cite{tanaka13a} followed a complementary approach by using a new line list for r-process 
elements based on the VALD data base \citep{kupka00} and applied it in radiative transfer 
calculations to a 2D, axisymmetric ejecta model that is based on the merger simulations of
\cite{hotokezaka13a}. They also favour opacity values around 10~cm$^2$/g.

In this study we make use of the 3D ejecta geometry as calculated in
hydrodynamic simulations that include the heating from radioactive decays. The overall hydrodynamic 
evolution with a particular focus on the effect of the r-process heating is discussed 
in a companion paper, thereafter referred to as `Paper~I'  \citep{rosswog13c}. 
Implications of these  calculations for light curves of accompanying radio flares \citep{NP11,piran13a} will be 
discussed elsewhere. 
The rest of this paper is structured as follows. In Sec.~\ref{sec:hydro}, we describe our 
hydrodynamics results and in Sec.~\ref{sec:nucleo}, we briefly review the nucleosynthesis in the 
ejecta. We discuss the nuclear energy release and the expected opacities, mainly based
on the recent work by \cite{kasen13a}. Again for some parts we will refer to paper
I for more details. In Sec.~\ref{sec:Lightcurves}, we discuss in detail the radioactive
transients that are expected from the dynamic ejecta. Our results are based on the true 
matter distribution, but we also examine the accuracy of various semi-analytic approaches.
In Sec.~\ref{sec:nu_winds}, we discuss a second radioactive transient that results from
neutrino-driven winds. This radioactive material is also produced in an r-process,  
though different from the one that occurs inside the dynamic ejecta. While the latter 
is a `strong r-process' that very robustly produces in each event the same
abundance pattern of very heavy elements ($A>130$), the former is a `weak' r-process 
whose abundances vary depending on the detailed neutrino luminosities and which produces 
less heavy nuclei in a range $50 \la A \la 130$. This has implications for both 
the radioactive half-lives and the opacities and therefore the properties of this 
second radioactive transient differ from the more commonly considered radioactive 
transient that arises from the dynamic ejecta. 
In Sec.~\ref{sec:detect}, we discuss the detectability of these transients and we 
summarize and conclude in Sec.~\ref{sec:conclusion}, where we also 
briefly discuss the implications of
our results to the recent detection of an nIR transient accompanying GRB 130603B 
\citep{tanvir13,berger13}.

\section{Hydrodynamics} 
\label{sec:hydro}

We begin by briefly summarizing our hydrodynamic simulations, for more details we refer
to Paper~I and the references therein. The simulations of this paper begin from the final configurations of calculations
presented in \cite{rosswog13a} and \cite{rosswog13b}. 
These simulations were performed with a 3D smoothed particle hydrodynamic (SPH) code that has been described in 
previous publications \citep{rosswog00,rosswog02a,rosswog03a,rosswog07c}. For reviews
of the SPH method, we refer to \cite{monaghan05}, \cite{rosswog09b} and \cite{springel10}.
The subsequent long-term evolution is performed with a variant of the above code that uses
different units and physics ingredients \citep{rosswog08b}. In particular, we switch 
from the Shen equation of state \citep[EOS;][]{shen98a,shen98b} to the Helmholtz EOS 
\citep{timmes00a}, and when we hit the lower input limits of the latter, we smoothly 
switch over to a Maxwell--Boltzmann gas plus radiation.
We systematically account for the heating from radioactive decays by employing fit formulae
for the nuclear energy generation rate, $\dot{\epsilon}_{\rm nuc}$, and the average nucleon
and proton numbers which are needed to call the Helmholtz EOS. For the explicit forms of
the fit formulae, we refer to appendix A of Paper~I.\\
To follow the long-term evolution of the dynamically ejected matter, we cut out the central
object and accretion torus, typically at a radius of $R_{\rm cut}= 300$ km, and replace the 
corresponding matter with a point mass. We follow the evolution of the ejecta up to
100 years for a number of exemplary systems: a) an equal mass 
merger with $2 \times 1.4 \msun$, b) a merger with a slight asymmetry, 1.4 and 1.3 $\msun$, 
c) a merger of a 1.6--1.2 $\msun$ system and finally d) the merger of a 1.8 $\msun$ ns with a 
1.2 $\msun$ ns. The physical parameters of these simulations are summarized in
Table~\ref{tab:runs}.

\begin{figure*}
 \centerline{\includegraphics[width=0.9\textwidth]{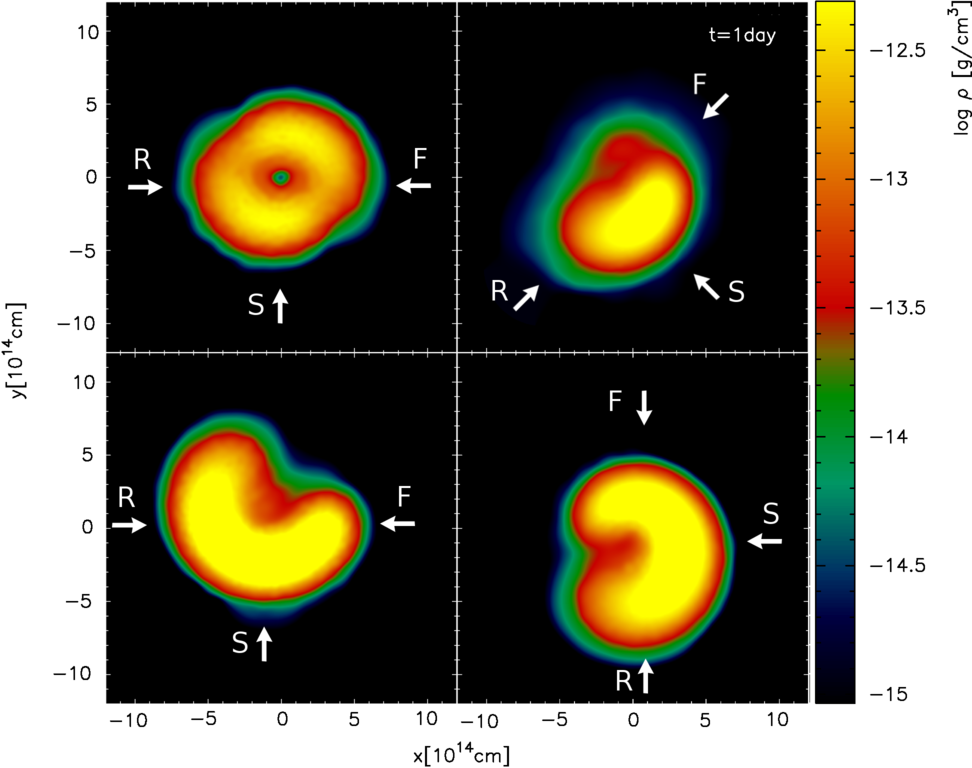}}
\caption{
    Density in the orbital plane at one day after the merger, for the
  different merger cases.
  Top left: 1.4--1.4~$\msun$, top-right: 1.3--1.4~$\msun$, bottom-left : 
  1.6--1.2~$\msun$ and bottom right:  1.8--1.2~$\msun$.
  The arrows indicate three different viewing angles that are referred to in
  later figures, and the fourth `top' view direction (not shown) is always
  perpendicular to the orbital plane.
} 
\label{fig:density}
\end{figure*}

In Fig.~\ref{fig:density}, we show the density distribution in the orbital plane at
$t=1$ d after the merger which is characteristic for the peak time of macronovae.
In the equal-mass case (top left), the remnant is symmetric with two tidal tails, 
for the other cases only one tidal emerges which becomes more pronounced
with increasing deviations of the mass ratio from unity.

\begin{figure*}
 \centerline{
 \begin{tabular}{cc}
  \includegraphics[width=0.45\textwidth]{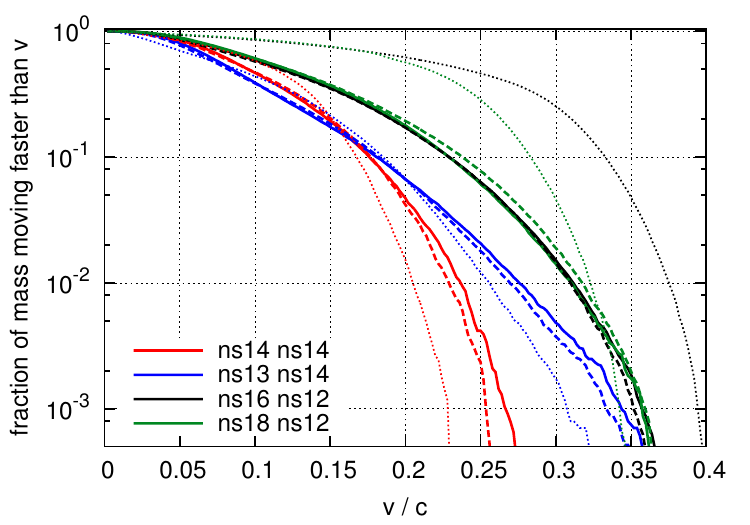}  &
  \includegraphics[width=0.45\textwidth]{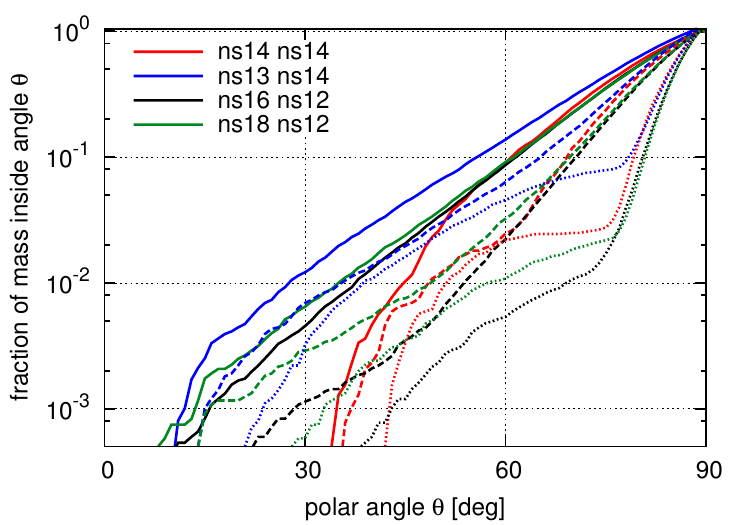}
  \\
  (a) & (b)
 \end{tabular}
 }
\caption{ Snapshots of radial and angular mass distributions for different merger
 cases at different times. Shown are initial distributions right after the merger
 (dotted lines), distributions at 2~s (dashed lines) and at late time
 (solid lines):
 (a) spherical mass distribution, mass with velocity larger than $v$ as a function
 of  $v$. One can clearly see the acceleration of the matter due to the
 radioactive decay. The distribution settles to a simple homologous expansion
 after a few hours.
 (b) Mass within angle smaller than $\theta$ as a function of polar angle
 $\theta$. This plot demonstrates that there is no significant mass in the
 polar region even after the remnant has puffed up due to radioactive heating. 
 Thus, the dynamic ejecta leave a rather empty wide funnel along the rotation
 axis in which the neutrino-driven wind can escape.
}
\label{fig:mass_distribution}
\end{figure*}

Fig.~\ref{fig:mass_distribution} shows snapshots of the radial and polar angle
mass distributions at different epochs. Mass accelerates due to radioactive
heating, but settles quickly to a homologous expansion. While there is
significant acceleration from the moment of merger until after the neutron
freeze-out (dotted and dashed lines in Fig.~\ref{fig:mass_distribution}), from
there on distributions change only slightly. 
The cases of 1.6--1.2 and 1.8--1.2~$\msun$ have much broader initial
radial distributions, while at later times the distribution shrinks towards
the centre.
This does not mean, however, that the mass is decelerating, because the radial
distributions in Fig.~\ref{fig:mass_distribution}a are calculated with respect
to the density maximum.
If the location of the latter changes with time, so does the distribution. 
In both merger cases mentioned above the tidal tails are initially highly non-spherical
with a compact dense core and an extended tail. 
As the merger remnant expands and puffs up due to radioactive heating, the
density maximum moves closer to the geometric centre and the distribution
becomes more centrally condensed.

Fig.~\ref{fig:mass_distribution}b shows  the amount of mass contained in a
cone with an opening angle $\theta$ with respect to the initial binary rotation axis.
This quantity is important since it determines whether the neutrino-driven wind that we discuss in 
\S \ref{sec:nu_winds} can escape along the rotation axis or not.  This figure shows that  only a 
small amount of mass is contained in the polar region. This implies  that a neutrino-driven
wind that is produced shortly after the merger can likely escape without a significant 
interaction with the dynamic ejecta. 

As discussed in Paper I, all of our models show homologous expansion with the degree of 
homology up to 0.01~per cent at the time period which is relevant for the macronova calculations 
(1~d).
 

\begin{table}
  \caption{
    Overview of the performed simulations; all neutron stars have zero initial spin. 
    $N_{\rm SPH}$ is the number of SPH particles in the initial ns$^2$ simulation,
    and $t_{\rm end}$ is the time when the initial ns$^2$ merger simulation is stopped;
    the hydrodynamic evolution of the ejecta is subsequently followed up to a time
    of 100 years. $m_{{\rm ej},-2}$ is the ejected mass in 
    units of $10^{-2}\msun$, and $v_{1/2}$ is the median velocity in
    the final radial mass-velocity profile (see Fig.~\ref{fig:mass_distribution}a).
  } 
  \label{tab:runs}
  \vspace*{0.3cm}
  \begin{tabular}{@{}clllll@{}}
  \hline
   Run ($m_1-m_2$) & 
   \hspace{-1mm}$N_{\rm SPH}$ &
   \hspace{-3mm}$t_{\rm end}$ (ms) &
   \hspace{-1mm}$m_{{\rm ej},-2}$  &
   \hspace{-1mm}$v_{1/2} (c)$\\
   \hline \\  
   A $(1.4-1.4)$ & $1.0\times 10^6$ &  13.4  & $1.3 $ &  0.095\\
   B $(1.3-1.4)$ & $2.7\times 10^6$ &  20.3  & $1.4 $ &  0.086\\  
   C $(1.6-1.2)$ & $1.0\times 10^6$ &  14.8  & $3.3 $ &  0.119\\
   D $(1.8-1.2)$ & $1.0\times 10^6$ &  21.4  & $3.4 $ &  0.121\\
  \end{tabular}
\end{table}

\section{Nucleosynthesis and opacities}
\label{sec:nucleo}

The macronova transients that we are interested in are caused by radioactive matter
that is ejected into space during a compact binary merger. There are at least three
ejection channels for neutron-rich matter in a compact binary merger: a) the dynamic ejecta that are launched
by hydrodynamic interaction and/or gravitational torques 
\citep{rosswog99,bauswein13a,hotokezaka13a,rosswog13b}, 
b) neutrino-driven winds from the remnant 
\citep{ruffert97a,rosswog02b,rosswog03b,rosswog03c,surman08,dessart09,caballero12,wanajo12}
and c) the final disintegration of accretion discs that occurs when viscous dissipation 
and the recombination of nucleons into  light nuclei conspire to unbind a 
substantial fraction of the initial accretion disc 
\citep{lee07,beloborodov08,metzger08,lee09,metzger09b,fernandez13}.

Each of these channels may eject matter with different properties which impacts 
on both nucleosynthesis and possible radioactive transients. So far, the dynamic
ejecta have received most attention although the other channels may have interesting 
consequences as well. The dynamic ejecta are launched essentially immediately
at first contact,  even before the 
neutrino emission has reached substantial values.
Therefore, the ejecta electron fraction is very close to the initial cold 
$\beta$-equilibrium value of around $Y_e= 0.03$ \citep{korobkin12a}. Under these
conditions matter undergoes a `strong' r-process and produces very heavy nuclei
($A > 130$) with a robust abundance pattern that is independent of the 
parameters of the merging compact binary \citep[][Paper I]{korobkin12a,bauswein13a}. 

The neutrino-driven wind case is different in several respects. Here, matter is 
not ejected immediately but instead exposed to neutrino fluxes for a much longer 
time and it is only gradually accelerated by capturing (anti-)neutrinos in $\beta$-processes. 
Therefore, it is the relative neutrino luminosities and energies that determine 
the final $Y_e$ values, like in the case of proto-neutron stars \citep{qian96b}. 
The differences in the neutrino properties of our four considered cases,
see table~1 in Paper~I, therefore lead to a relatively large spread in the asymptotic 
wind electron fractions: 0.28 for run A, 0.30 for run B, 0.36 for run C and 0.40
for run D. In this regime, the nucleosynthesis is actually rather sensitive to the exact 
value of the electron fraction; therefore, the four merger cases produce 
a substantial spread in the resulting abundance patterns. In all cases, our
model, see Paper~I for more details, produces abundance 
distributions in the range $50 \la A \la 130$ that show substantial 
variations from case to case due to the different initial $Y_e$. Our wind model is 
for sure very simple; it is mainly meant to illustrate some basic points and 
qualitative features: a) a neutrino-driven wind produces also r-process, though 
only lower mass, `weak' r-process contributions and b) it can plausibly produce 
another transient EM event that is powered by radioactivity, though not $^{56}$Ni. 

For the third channel, the dissolution of accretion discs on viscous time-scales,
recent models \citep{fernandez13} find that also $\sim$10~per cent could become unbound 
with electron fractions of $\sim 0.2$. 
To our knowledge, no detailed
nucleosynthesis results exist yet for this material, although
\cite{fernandez13} argue for this outflow channel to also produce heavy
r-process elements \citep[based on the parametrized study of][]{hoffman97}.
In the following, we will focus entirely on the former two channels.

The spread in the abundance patterns also translates to the opacities.
\cite{kasen13a} suggest that the elements with half-filled f-shells such as
lanthanides and actinides should have very high opacities around 10 cm$^2$/g,
consistent with the conclusions of \cite{tanaka13a}. We therefore adopt
this value in our study. For the wind material, we adopt a lower value, $\kappa=1$ cm$^2$/g.
This needs admittedly further study in the future, but we consider this justified
at the current stage since our model is mainly to illustrate qualitatively the 
plausibility of a second transient. 
It is worthwhile pointing out, however, that for these grey opacities our results satisfy simple
scaling laws and the results apply, after this scaling, to any system in which  the opacity 
is frequency independent.

We had recently found \citep{korobkin12a}  that the nuclear energy generation
rate is relatively independent of details and has a simple temporal dependence:
it is constant within roughly the first second and subsequently decays according to
a power law $\propto t^{-\alpha}$, where $\alpha= 1.3$, 
similar to the results found by other groups~\citep{metzger10b,goriely11a,roberts11}.
The energy generation and the average nucleon and proton number that are needed
to call the EOS can be easily fit as a function of time and they are included
into the hydrodynamic simulation via fit formulae; the explicit expressions are given in 
appendix~A of Paper~I.

All the nucleosynthesis results that we are quoting here and in Paper~I have 
been obtained with the nuclear reaction network of Winteler 
\citep{winteler12,winteler12b} which is an update of the BasNet network 
\citep{thielemann11}.

\section{The light curves}
\label{sec:Lightcurves}

We did not carry out detailed radiation transfer simulations, instead we use simple
approximations, discussed below, that capture the main characteristics of the radiation
from the expanding remnant.  The basic idea is that photons can escape in a dynamical 
time from the region  $R>R_{\rm diff}$, where $R_{\rm diff}$ is the radius from which the 
diffusion time,  $t_{\rm diff}$,  equals the dynamical time, $t_{\rm dyn}=r/v=t$. These photons 
diffuse out in the optically thick regime until they reach the photosphere at
$\tau=2/3$. The photons escape from the photosphere with a quasi-thermal, possibly 
blackbody but basically unknown spectrum. The details of this spectrum depend very strongly 
on unknown features of the opacity.  Emission generated at the region  $\tau< 2/3$ 
escapes most likely at high energies and  with an unknown spectrum.  We
neglect this radiation in our calculations as it is most unlikely that it will
be at the IR--optical range that we consider here.
Indeed, all X-rays and $\gamma$-radiation produced directly in
radioactive decays freely escape, while high-energy electrons need several
scattering events before they could induce bound--bound transitions in optical
or IR. But in the low-density transparent medium, they are not able to
efficiently thermalize, so their energy will be ultimately lost to expansion.
The luminosity of this optically thin component is shown as blue dotted 
line in Fig.~\ref{fig:analytic_lum}.

We begin in \S \ref{sec:analytic} with simple analytic estimates of the escaping emission
and  continue in \S \ref{sec:rad_methods} with a detailed description of how we numerically 
calculate the radiation from the simulation results. In \S \ref{sec:results}, we describe our
results from the 3D matter distribution.

\subsection{Analytic estimates}
\label{sec:analytic} 
Before considering the numerical solutions we consider a few analytic 
approximations. The basic idea behind these approximations is that 
the radiation diffuses out from the region in which  
$t_{\rm diff}\le t_{\rm dyn}$, i.e. the 
diffusion time is less than or equal to the dynamical time. 
The latter evolves uniformly and therefore simply equals $t$.
The  approximations differ in the estimates of the diffusion time. 

We consider a spherical homologous system and define $m(v)$, the mass with velocity 
larger than $v$, which is shown in Fig.~\ref{fig:mass_distribution}a. Since the system is 
highly homologous, this mass is located further out relative to the mass that is moving at lower 
velocities. The maximal velocity is denoted by $v_{\rm max}$.

\begin{figure}
 \centerline{ \includegraphics{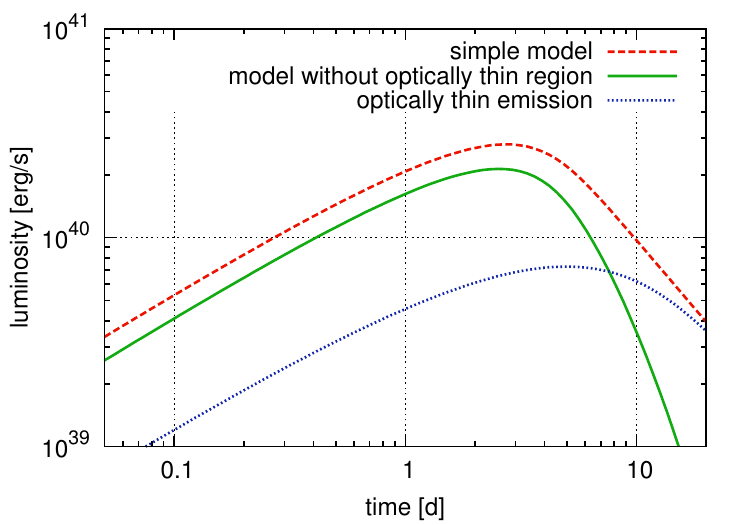}}
\caption{
  A comparison of the two approximate bolometric light curves given 
  by two approximations for a fiducial profile having mass $m_{\rm ej} = 0.01\msun$
  and density distribution $\rho(v) = \rho_0 (1-v^2/v_{\rm max}^2)^3$, where
  $v_{\rm max}=0.35\,c$.
  The opacity is taken to be $\kappa=10\,{\rm cm}^2/{\rm g}$. 
  The simple model is described by a dashed line while the one with the
  optically thin region excluded is shown by a solid line.
  In the second approximation the effectively contributing radiating layer is 
  smaller, leading to a slightly earlier and less luminous signal, which also has
  a much steeper decline.
  The blue dotted line shows the difference between the two, which corresponds
  to the power emitted in high-energy photons by the optically thin layer.
} 
\label{fig:analytic_lum}
\end{figure}

The simplest and most commonly used approximation estimates the diffusion time as 
(e.g. \cite{piran13a})
\begin{equation}
t_{\rm diff} = 
  {{\tau (v_{\rm max}-v)t}\over{ c}}= 
  {{m \kappa  }\over{ 4 \pi c v t   } } ,
\end{equation}
where the optical depth for the mass $m(v)$ is approximated as if this mass is 
distributed uniformly within a spherical layer of radius $vt$ and thickness
$(v_{\rm max}-v)t$.
This leads to an implicit equation for $v$:
\begin{equation}
{{m (v)   }\over{  v   } }  = {4 \pi  c t^2 \over \kappa}. 
\end{equation}
In the first approximation, neglecting the fact that emission from an optically 
thin region ($\tau < 2/3$) escapes freely and not necessarily in the IR--optical 
or UV bands, we estimate the bolometric luminosity as:
\begin{equation}
\label{eq:lum}
L (t) = \dot{\epsilon}(t) m(v) 
      = \dot{\epsilon}_0 (t/t_0)^{-\alpha} m(v) , 
\end{equation} 
where $t_0$ denotes the onset of the power law decay ($\approx 1$ s),
and $\dot{\epsilon}_0$ is the heating rate at $t=t_0$.
This approximation is commonly used to estimate the peak time, $\tilde t_{\rm p}$, 
and peak bolometric luminosity, $\tilde L_{\rm p}$: 
\begin{eqnarray}
\label{eq:peakt}
\tilde t_{\rm p} &\approx& \sqrt{ {\kappa m_{\rm ej} \over 4 \pi c \bar v}} = 
  4.9 \; {\rm d} \;
     \left(\frac{\kappa_{10}m_{\rm ej,-2}}{\bar v_{-1}}\right)^{1/2}
,
\end{eqnarray}
\begin{eqnarray}
\label{eq:peakL}
\tilde L_{\rm p} &\approx& \dot{\epsilon}_0 m_{\rm ej} \left({ 
\kappa m_{\rm ej} \over 4 \pi c \bar v t_0^2} \right )^{-\alpha/2} = 
\nonumber \\
 &=& 2.5\times10^{40}\,\frac{\rm erg}{\rm s}\;
           \left(\bar v_{-1}\over\kappa_{10}\right)^{\alpha/2}
           m_{\rm ej,-2}^{1-\alpha/2},
\end{eqnarray}
where we defined
  $\kappa_{10} = (\kappa/10\,{\rm cm}^2{\rm g}^{-1})$,
  $m_{\rm ej,-2} = (m_{\rm ej}/0.01\,\msun)$ and
  $\bar v_{-1} = (\bar v/0.1\,c)$.
It is valid  in the intermediate regime where radiation is produced by a
significant fraction of the overall mass. If we are bold enough, we can even
estimate the effective temperature. Assuming that the effective emitting area
is of the order of $4\pi (v t)^2$, we obtain: 
\begin{eqnarray}
\label{eq:peakT}
\tilde T_{\rm eff} &\approx&
 \left( {\dot{\epsilon}_0 c \over \sigma_{\rm SB}} \right)^{1/4}
 \left({ m_{\rm ej} \over 4 \pi c t_0} \right )^{-\alpha/8}
 \kappa^{-(\alpha+2)/8}
 \bar{v}^{(\alpha-2)/8}
\nonumber \\
 &=& 2200\,{\rm K}\;
 \kappa_{10}^{-(\alpha+2)/8}
 \bar v_{-1}^{(\alpha-2)/8}
 m_{\rm ej,-2}^{-\alpha/8}.
\end{eqnarray}
In the second approximation we subtract the mass $m(v_{2/3})$ contained in the
optically thin region:
\begin{equation}
\label{eq:lum1}
L (t) = \dot{\epsilon}(t) (m(v) - m(v_{2/3})).
\end{equation} 
Fig. \ref{fig:analytic_lum} depicts a comparison of the resulting bolometric light 
curves with both approximations for a  mass distribution of the form 
$\rho(v) = \rho_0 (1- v^2/v_{\rm max}^2)^3$. 
The simple model gives more or less the time at which that the whole system begins to 
emit and indeed shortly after this peak the emission settles to the late phase 
of $t^{-\alpha}$ decay. The model with the optically thin region excluded
produces a luminosity which is smaller by about a factor of $\sim1.2$ in the raising
phase and a more rapid decline ($\propto t^{-(3+\alpha)}$), reflecting the fact that
increasingly smaller fractions of emitted energy can be reprocessed into the
UV, optical or infrared bands.
The rest of the nuclear power produced in the optically thin region is lost to
high-energy emission with an unknown spectrum. This power is shown in
with the blue dotted line in Fig.~\ref{fig:analytic_lum}. 

\subsection{Numerical treatment}
\label{sec:rad_methods}

\begin{figure*}
 \centerline{
 \begin{tabular}{cc}
 \includegraphics[height=0.330\textheight]{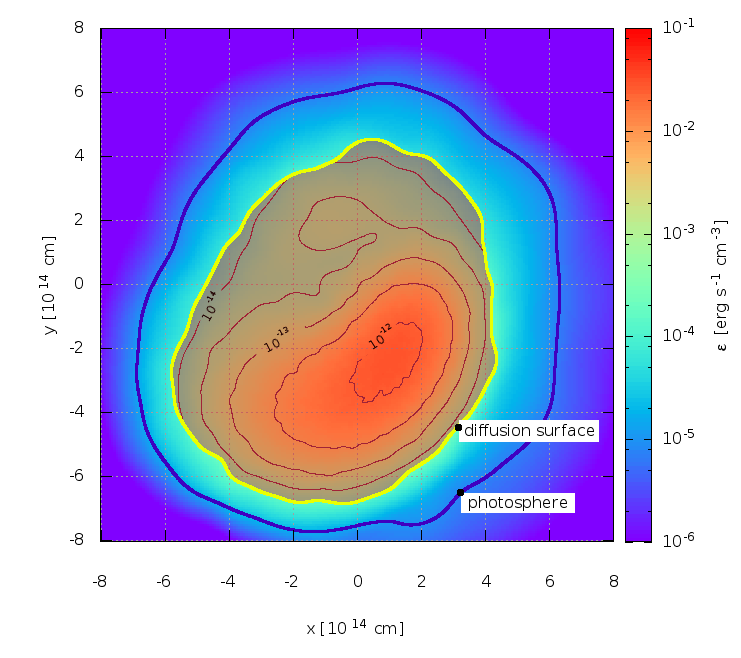} &
 \includegraphics[height=0.335\textheight]{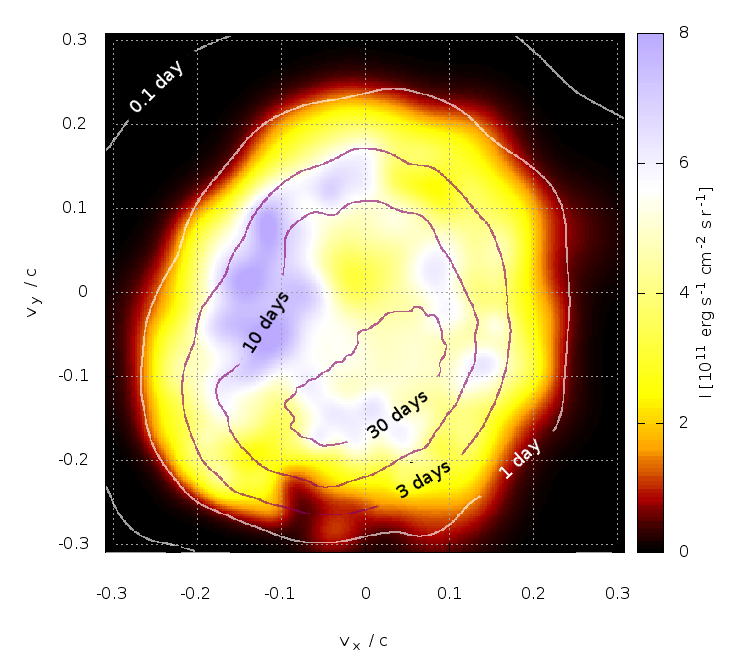}    \\
  (a) & (b)
 \end{tabular}
 }
\caption{ 
  Configuration of the remnant for the merger case 1.3--1.4~$\msun$ at
  $t_0=1$~d:
  (a)  
    $XY$-cut through the density distribution. The thick blue and yellow lines
    delineate the photosphere and the diffusion surface respectively, and the thin
    red lines inside the diffusion surface are for five density isocontours,
    equally spaced (in log scale) from $10^{-14}$ to $10^{-12}$~g/cm$^3$.
    Colour-coded is the volume emissivity (nuclear energy generation rate per
    volume).  The blue layer between the diffusion surface and the photosphere
    is the main contributor to the total luminosity of the remnant. Photons
    inside the diffusion surface are `trapped' due to high optical depth,
    while the photons outside photosphere have a very hard spectrum and escape
    before their energy is converted to visible or infrared.
  (b)
    Map of specific intensity, emitted from the photosphere in the
    $z$-direction (in the direction of the `top' observer),
    plotted in velocity space. Visible shape of the photosphere and the column
    density determine the relative brightness of the transient in a given
    direction. Superimposed are the cuts of the photosphere in the $XY$ plane at
    different times, which also coincide with the isocontours of optical depth.
} 
\label{fig:photospheric_1}
\end{figure*}

For our numerical models we can produce light curves from a sequence of 3D 
snapshots of the expanding material. Alternatively, one can make use of the fact
that the expansion is highly homologous for $t>0.05$~d, 
as shown in fig.~7 of Paper I.
Therefore, most of the dynamical quantities can be 
obtained by a simple rescaling of a reference snapshot, taken for example at $t_0=1$~d.
Such a reference snapshot from run B (1.3--1.4~$\msun$) is shown in Fig.~\ref{fig:photospheric_1}. 
One can distinguish three regions: the optically thick
inner region, a `radiating volume' between optical depths of $\tau_{\rm diff}$
and 2/3 and finally the transparent envelope with $\tau < 2/3$. In the inner region, 
the density is high and the photons are efficiently trapped inside
the diffusion surface, defined by the condition that the photon diffusion time 
$t_{\rm diff}$ equals the dynamical time $t_{\rm dyn}$. The former can be estimated 
as $t_{\rm diff} = \zeta\tau/c$, where $\zeta$ is the length of the shortest photon 
`escape route', normally following the negative gradient of the optical depth.
The inner region does not contribute to the observed luminosity since the 
photons are trapped. The outer region likely does not contribute either,
because the photons escape with large energies determined by the
nuclear reactions.

We are therefore mainly interested in the region between the diffusion surface and
the photosphere, with the latter defined by the condition $\tau=2/3$.
For general configurations, both of these surfaces can have non-trivial shape.
Fig.~\ref{fig:photospheric_1}a shows the contours of the photosphere and the
diffusion surface at $t_0=1$~d, along with the distribution of emissivity in
the $XY$ plane. We estimate the luminosity by integrating the emissivity over
the `radiating volume' between the photosphere and the diffusion surface (see
Fig.~\ref{fig:photospheric_1}a).

\begin{figure}
 \centerline{
 \includegraphics[width=0.5\textwidth]{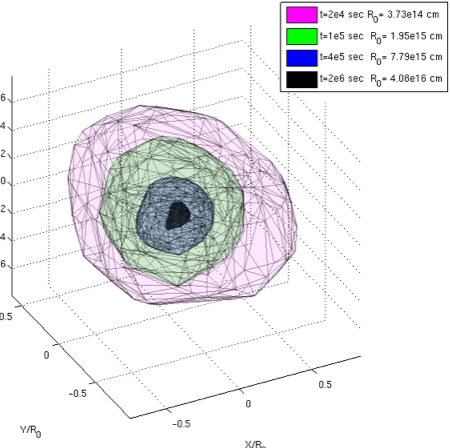}}

\caption{ The photospheric  surfaces at different moments of time ($2\time 10^4$, 
$10^5$, $4 \times 10^5$ and $2 \times 10^6$~s). The scales are normalized and 
should be multiplied by $3.73 \times 10^{14}$, $1.95 \times 10^{15}$, 
$7.79 \times 10^{15}$ and $4.08 \times 10^{16}$ cm, respectively.  These surfaces 
reflect the observed shape of the macronova. }
\label{fig:photospheric_2}
\end{figure}

Due to the homologous expansion, we can use the particle velocities as 
time-independent coordinates. Then, the photosphere and the diffusion surface
can be found as level surfaces of the optical depth  $\tau(\v)$ and
$\tau(\v)\zeta(\v)$, specifically the surfaces defined by:
\begin{equation}
\tau(\v) = \frac23 \left(\frac{t}{t_0}\right)^{-2},\qquad
\tau(\v)\zeta(\v) = c\,t\,\left(\frac{t}{t_0}\right)^{-2}.
\end{equation}
Fig.~\ref{fig:photospheric_1}b shows the contours of $\tau(\v)$ in velocity
space for several successive times, and Fig.~\ref{fig:photospheric_2} shows
these contours in 3D. The photosphere traverses a sequence of
embedded level surfaces of $\tau(\v)$, and so does the diffusion surface for
the level surfaces of $\tau(\v)\zeta(\v)$.
The combined effects of these two surfaces determine the luminosity and the
effective temperature of the escaping emission. 

We identify $\tau(\r)$ and $\zeta(\r)$ at $t_0=1$~d
and perform the volume integration with properly rescaled quantities. 
With our assumption of constant opacity, the optical depth can be
calculated by integrating ${\rm d}\tau=\kappa \rho(z) {\rm d}z$ along the path following
the negative density gradient.
Since the density is given in SPH representation, it is convenient to
express $\tau(\r)$ in the same way:
\begin{equation}
  \tau(\r) = \sum_b \tau_b \frac{m_b}{\rho_b}\; W_b(\r),
  \label{eq:tau_SPH}
\end{equation}
where $W_b(\r)= W(|\r-\r_b|,h_b)$ is a smoothing kernel, 
$h_b$ is the SPH particle smoothing length, and $m_b$ and $\rho_b$ are its
mass and density. It suffices to calculate $\tau_b$ for every particle to
properly approximate optical depth everywhere, and the similar SPH expansion
is employed to represent $\zeta(\r)$.

The total luminosity is calculated as:
\begin{equation}
  L = \int_{\tau(\r)>2/3}^{\tau(\r) < c t/\zeta(\r)} 
             \dot{\epsilon}(t)\rho(\r) {\rm d}^3\r
    \approx \sum_{\tau_b>2/3}^{\tau_b < c t/\zeta_b}
             \dot{\epsilon}(t)m_b,
\label{eq:lum2}             
\end{equation}
where the final sum runs over all particles with $\tau_b>2/3$ and
$\tau_b < c t/\zeta_b$.

To compute the luminosity radiated in a given direction, in principle one needs
to solve complete radiative transfer equations. The energy originating deep in
the bulk of the radiation layer will be transferred to the photosphere and
spread over a finite region.
To mimic this effect, we adopt the following simple approach.
Because the luminous flux is more likely to stream in the direction of
negative gradient of optical depth, the directional luminosity can be
approximated by:
\begin{eqnarray}
  \frac{dL}{d{\bf\Omega}}(\k)
  &= \int_{\tau(\r)>2/3,\k\cdot\n>0}^{\tau(\r) < c t/\zeta(\r)} 
             \k\cdot\n \;\dot{\epsilon}(t)\rho(\r) d^3\r \approx
  \\
  &\approx \sum_{\tau_b>2/3,\k\cdot\n>0}^{\tau_b < c t/\zeta_b}
             \k\cdot\n_b\;\dot{\epsilon}(t)m_b,
\end{eqnarray}
which differs from the total luminosity (\ref{eq:lum2}) by a factor
$\k\cdot\n$, representing the angle between the observer and the unit vector
in the direction of negative gradient of $\tau$.

\begin{figure}
\includegraphics[width=0.5\textwidth]{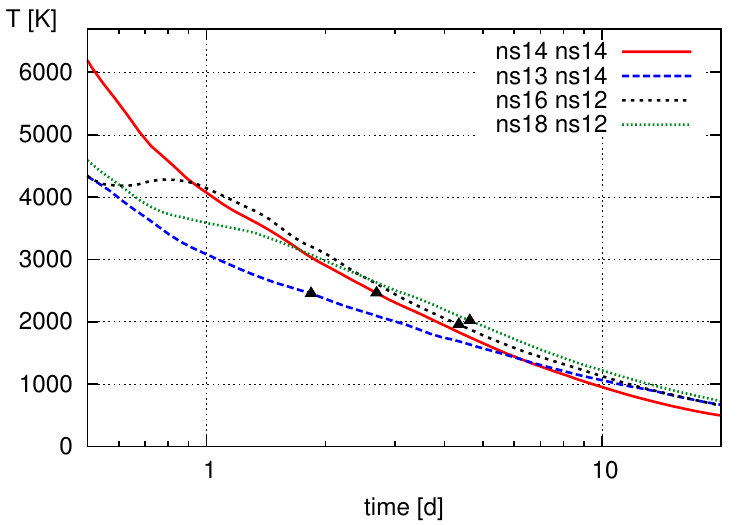}
\caption{The effective photospheric temperature for the four different merger
  models. These temperatures should be taken with caution as they do not
  reflect possible effects of the opacity. 
  The black triangles represent peak times of the bolometric lightcurves.
}
\label{fig:temperature}
\end{figure}

The averaged effective temperature presented in Fig.~\ref{fig:temperature} is simply
$T_{\rm eff} = \left({L}/{\sigma_{\rm SB}S_{\rm ph}}\right)^{\frac14}$.
Here $S_{\rm ph}$ is the surface area of the photosphere, which is computed
using the following expansion in terms of SPH quantities:
\begin{eqnarray}
\int_\Sigma dS 
  &=& \sum_b \frac{m_b}{\rho_b} \int_\Sigma W_b(\mathbf{r}) dS 
\\
  &\approx& \sum_{b\;\cap\;\Sigma} 
             \frac{m_b}{\rho_b} 
             h_b\;\sigma_b(h_b,\Delta_b/h_b),
\end{eqnarray}
where $h_b \sigma_b(h_b,\Delta_b/h_b)$ is the contribution from the 
smoothing kernel of $b$ with the surface integrals replaced by integrals over
respective tangent planes. 
The quantity $\Delta_b$ for a particle $b$ is its distance from the
integration surface.
For the cubic spline kernel that we use~\citep[see, e.g.][]{schoenberg46},
\begin{eqnarray}
\sigma_b(h, q) = 
\frac{1}{h^2}\;\left\{
  \begin{array}{ll}
  0.7 - q^2 + 0.75q^4 - 0.3q^5, & {\rm if}\; q<1,
  \\
  0.8 - 2q^2 + 2q^3   & 
  \\
  \qquad \qquad- 0.75q^4 + 0.1q^5, & {\rm if}\; 1\leq q<2,
  \\
  0, & {\rm if}\; q\geq2.
  \end{array}
\right.
\end{eqnarray}
The `kernel slices' defined in this manner are normalized such that
$ 2 h^2 \int_0^{2}\sigma_b(h, q)dq = 1$.

\subsection{Results from the 3D remnant matter distribution}
\label{sec:results}

We calculate the light curves following the method outlined in the previous section 
for a sequence of late time (hours and days) distributions of the expanding 
dynamic ejecta. Here we deal exclusively with dynamic ejecta, while the additional 
signature from neutrino-driven winds is discussed later in \S \ref{sec:nu_winds}.

The dynamically ejected mass is around 0.01~$\msun$ and it has typical initial 
velocities of 0.1--0.2~$c$, see Table~\ref{tab:runs}. As discussed in \S \ref{sec:nucleo}
and in more detail in Paper~I, this matter consists of  a `strong' r-process material
with $A>130$ and likely has  large effective opacities around $\kappa=10$ cm$^2$/g  
\citep{kasen13a,tanaka13a}.  We use this value in all our
calculations  and all results are presented with this opacity. However, the
time in the problem scales like $\kappa^{1/2}$ and the peak luminosity scales
like $\kappa^{-\alpha/2}$, see equations~(\ref{eq:peakt}) and (\ref{eq:peakL}).
Therefore, the results can be easily rescaled to a configuration with a
different grey constant opacity.

\begin{figure}
 \centerline{ \includegraphics[width=9.5cm]{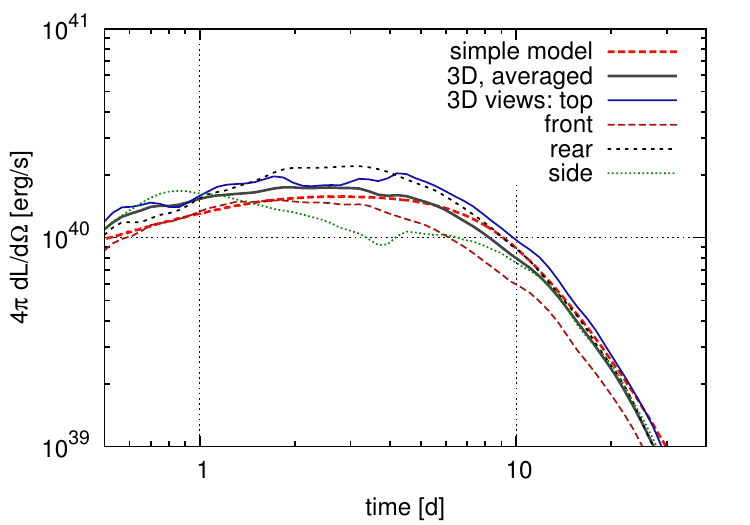}} 
 \caption{Bolometric equivalents for the 1.3--1.4~$\msun$ macronova, produced using
  different models. The thick dashed (red) line corresponds to the simple
  spherical model without $\tau<2/3$ region (see \S~\ref{sec:analytic}),
  calculated using the actual mass-velocity distribution $m(v)$ of the
  1.3--1.4~$\msun$ remnant.
  Different thin lines show the light curves calculated in 3D for four
  different observer views (see Fig.~\ref{fig:density}), and the thick black
  line represents the averaged light curve, which also equals the total luminosity.
 } 
\label{fig:bol_light_curve_1314}
\end{figure}

The properties of the macronova light curve depend mainly on four factors: 
the total mass ejected, its composition which determines the opacity, 
the mass--velocity distribution and the spatial distribution of mass.
Fig.~\ref{fig:bol_light_curve_1314} depicts the observed bolometric light
curve for the 1.3--1.4~$\msun$ macronova. In this case 0.014~$\msun$ were
ejected with a median velocity (after being accelerated by the radioactive
decay) of 0.086~$c$. The simplest order of magnitude estimates
(equations~\ref{eq:peakt} and \ref{eq:peakL}) give for  $m_{\rm ej} = 0.014 \msun$ and
$\bar{v}=0.086$~$c$ a peak time of $\tilde t_{\rm p} \sim 6 \times 10^5$~s 
$\sim 7$~d and a peak luminosity of $\tilde L_{\rm p} \sim  10^{41}$~erg/s. 
The peak time given by equation~(\ref{eq:peakt}) is slightly later than the one
obtained by the detailed calculations. However, the light curve is very flat
and the difference as compared with the simple estimate is moderate. 
This is consistent with the expectations for a non-spherical system where
the photospheric surface is larger, more radiation can leak out and the
peak flux is reached earlier.  Remarkably, the simple order of magnitude
estimate gives an effective temperature of 2200~K, which is very close to
the more detailed calculations. A similar comparison of the simple estimates
with the detailed calculation shows a similar trend for the other cases (see
Table~\ref{Tab:Results}).

\begin{table}
  \caption{Overview of the resulting light curves: the peak time, $t_{\rm p}$, peak
  luminosity luminosity, $L_{\rm p}$, and the effective temperature at the peak.
  Also shown are the simple estimates (equations~\ref{eq:peakt} and \ref{eq:peakL})
  of the peak time, $\tilde t_{\rm p}$, and the peak luminosity, $\tilde L_{\rm p}$.}
  \label{Tab:Results}
  \vspace*{0.3cm}
  \begin{tabular}{@{}clclcl@{}}
  \hline
   Run ($m_1-m_2$) & 
   \hspace{-2mm}$t_{\rm p}$(d)  & 
   \hspace{-2mm}$L_{\rm p}$ (erg/s) & 
   \hspace{-2mm}$T_{\rm eff}$(K) & 
   \hspace{-1.5mm}$\tilde t_{\rm p}$(d)  & 
   \hspace{-0.5mm}$\tilde L_{\rm p}$(erg/s)  \\
   \hline   
   A $(1.4-1.4)$ & 2.7 & $2.6\times10^{40}$ & 2500 & 5.7 & $2.7\times10^{40}$ \\
   B $(1.3-1.4)$ & 1.8 & $1.7\times10^{40}$ & 2500 & 6.2 & $2.6\times10^{40}$ \\  
   C $(1.6-1.2)$ & 4.3 & $4.4\times10^{40}$ & 2000 & 8.1 & $4.3\times10^{40}$ \\
   D $(1.8-1.2)$ & 4.6 & $3.9\times10^{40}$ & 2000 & 8.1 & $4.3\times10^{40}$ \\
   \hline   
  \end{tabular}
\end{table}

In Fig.~\ref{fig:bol_light_curve_1314}, we show the bolometric light curves for
four different viewing angles (viewing angles for all models are displayed in
Fig.~\ref{fig:density}). 
For comparison we show the average 3D bolometric light curve and the light curve
calculated for a spherical configuration with the same mass-velocity
distribution and the emission from the optically thin region being disregarded.
They all show a very broad peak ranging
from 1 to 10~d with a maximum at about 2~d. The maximal bolometric
luminosity is around $2\times10^{40}$~erg/s within a factor of 2 for
different viewing angles. The peak values of luminosity correspond to absolute
magnitudes of -11 to -12. 

 \begin{figure*}
 \centerline{ \includegraphics[width=0.9 \textwidth]{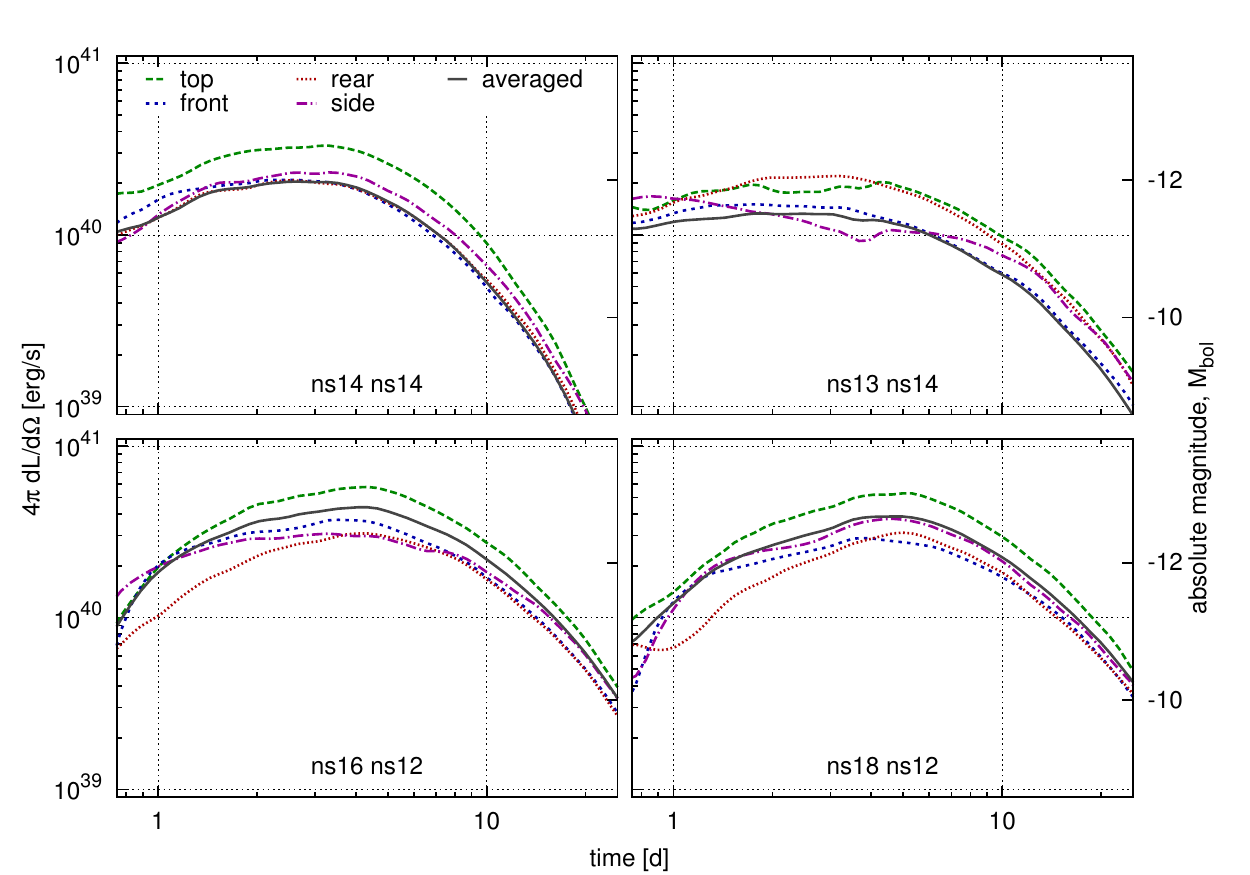}}
 \caption{Bolometric equivalents for all four merger models that we have
   examined, as seen by different observers (oriented as shown in
   Fig.~\ref{fig:density}). The black solid line shows the average light curve,
   which also corresponds to the total power emitted in UV, visible or
   infrared. The right axis is also labelled according to the corresponding
   absolute magnitudes.
 }
\label{fig:all_models}
\end{figure*}

Fig. \ref{fig:all_models} depicts the observed bolometric light curves for all
cases that we have examined. While the specific shape of the light curves
varies somewhat from one case to another, the overall trends remain the 
same with the emission from the top being typically brighter by about  a factor 
of 2 than the view from the front.  For the 
1.4--1.4~$\msun$ case, the peak is narrower, later (at around 2.5-3~d) and
brighter with absolute magnitude ranging from -12 to -12.5. For the two
remaining cases, 1.6--1.2~$\msun$ and 1.8--1.2~$\msun$ the peak is wide, late
(at around 4-5~d) and brighter with absolute magnitude ranging from -12.5
to -13.5. In these latter two cases, there is a clear shift of the peak as seen
by different observers.
Table~\ref{Tab:Results} summarizes the basic results of the different cases.
As expected from simple analytic scalings, the cases in which more mass is
ejected are more luminous and peak later. Also, the symmetric equal-mass case
1.4--1.4~$\msun$ exhibits a brighter and  later peak compared to the case of
1.3--1.4~$\msun$ with roughly the same mass. Not too surprisingly, the effective
temperatures are rather low.

\section{Neutrino driven winds} 
\label{sec:nu_winds}

\begin{figure*}
  \begin{center}
  \includegraphics[width=0.9\textwidth]{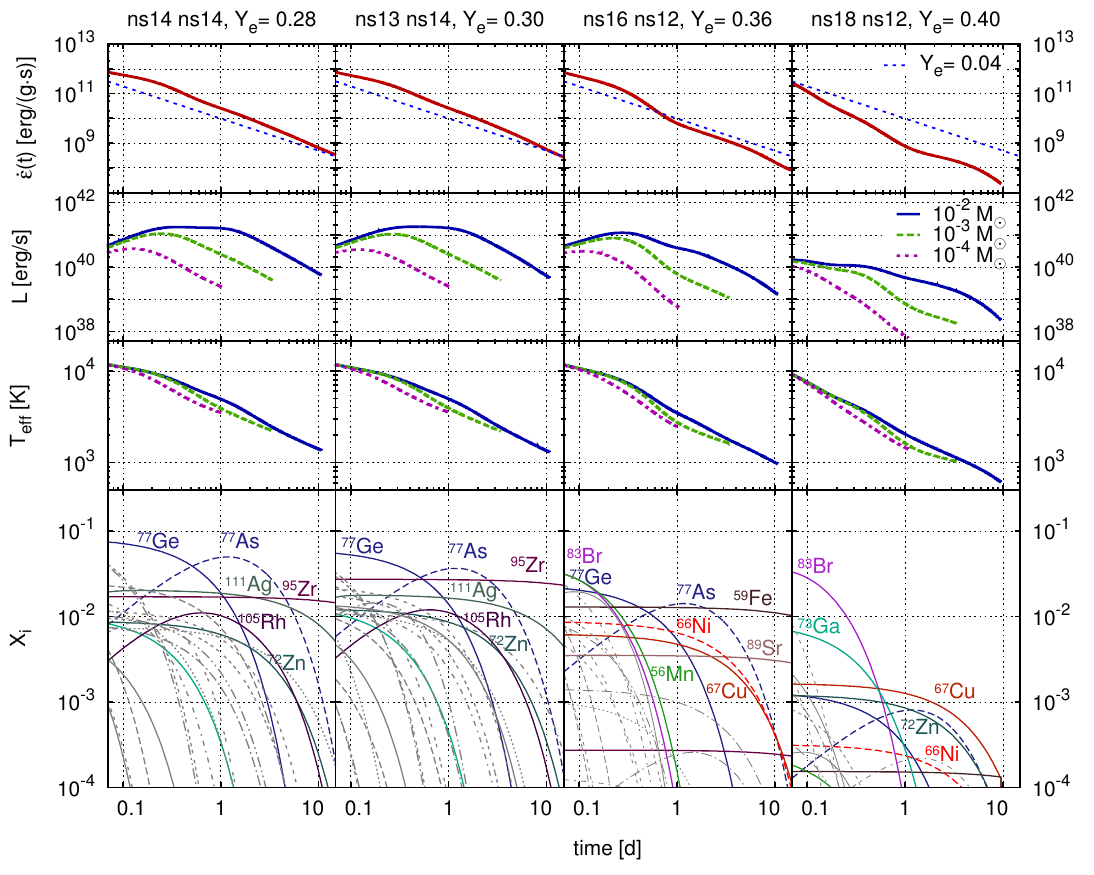}
  \caption{Radioactively powered transients due to neutrino-driven
            wind. Shown are the radioactive heating rates (first row), light
           curves (second row) and effective temperatures (third row) for all
           different runs. Note that these light curves are powered by
           different material than the `macronovae' from the dynamic ejecta.
           To highlight this difference, the top row also shows the nuclear
           heating power from the dynamic ejecta with $Y_e=0.04$.
           The bottom row displays the outflow composition in terms of the
           mass fractions of the dominant radioactive isotopes as a function
           of time. This material has also different opacities than the
           macronova material (we assumed $\kappa= 1$ cm$^2$/g).}
  \label{fig:transients_nu_winds}
  \end{center}
\end{figure*}

The neutrino-driven winds from an ns merger remnant can plausibly
complement the nucleosynthesis with `weak' r-process contributions ($50 \la A \la130$),
see section 4.2 of Paper~I and they can also plausibly produce an additional 
EM transient. \cite{barnes13a} discussed observational prospects 
for winds that are powered by $^{56}$Ni. However, as discussed in Paper I, 
we expect the $Y_e$ values in the wind to lie in a range from 0.28 to 0.40, i.e.
they are substantially below the $\approx0.5$ that is needed to synthesize
$^{56}$Ni. Nevertheless, we find  a range of radioactive isotopes with half-lives
long enough to power a short-lived transient, as can be seen in
Fig.\ref{fig:transients_nu_winds}, bottom row. 
The top row shows the corresponding nuclear heating rates compared to the
heating rates from the dynamical ejecta (dashed blue line on the plots). 
The larger variation due to individual isotopes, most pronounced in the
highest $Y_e$ case, is also apparent in the lightcurves, which could in
principle be used for identifying the isotope composition of the winds.
\\
As described in Paper~I, our model uses a simple linear expansion profile 
$\rho(t) = \rho_0 (1 + vt/R_0)^{-3}$ and starts from the entropy and $Y_e$ values
that are expected from the neutrino properties of our merger cases (8 $k_{\rm B}$/baryon;
$0.28 \le Y_e \le 0.40$).
Based on the results of \cite{dessart09}, their Fig. 2, we select the initial 
density $\rho_0=5\times10^7\;{\rm g}{\rm cm}^{-3}$, characteristic radius 
$R_0=200\;{\rm km}$ and expansion velocity $v = 0.11c$.
We parametrize the unknown wind mass in a range from $10^{-4}$ to
$10^{-2} \msun$, see the discussion in section~4.2 of Paper~I.
If the amount of mass ejected by the $\nu$-driven wind is too
small, the transient may be obscured from the observer by the previously
ejected dynamical component. 
However, the dynamical ejecta are mainly concentrated around the equatorial plane.
This can be seen in Fig.~\ref{fig:mass_distribution}b, which shows the fraction 
of the dynamically ejected mass within the polar angle $\theta$. 
Less than $10^{-4} \msun$ is obscuring the inner $30^o$ from the rotation axis, and
less than $10^{-3} \msun$ is within the inner $60^o$ for all our merger
simulations. The initial mass distribution is even flatter as can be seen from 
Fig.~\ref{fig:mass_distribution}b.
\\
To calculate the light curves from the radioactive wind material, we apply a simple spherical
semianalytic model (as described in Section~\ref{sec:analytic}) with the
radial density profile $\rho(r)=\rho_0(1-v^2/v_{\rm max}^2)^3$ with $v_{\rm
max}=0.35\;c$ and the opacity of $\kappa=1\;{\rm cm}^2\;{\rm g}^{-1}$.
As discussed earlier, such opacities are suitable for a mixture of elements heavier than iron but much
less opaque compared to lanthanides\footnote{Our nucleosynthesis calculations
indicate the presence of only trace amounts of lanthanides.}. 
Because of the low concentration of dynamical ejecta component in the polar
regions, we can also ignore possible admixing of lanthanides from the
dynamical component.
\\
We feed our model directly with the energy production rate taken  from the
nucleosynthesis network.
We find that the transients caused by neutrino-driven winds can reach up to $1.8\times 10^{41}$ erg/s
(for the most optimistic case with $10^{-2}  \msun$) and that they peak after 1 to 12~h in the UV/optical, 
see Table~\ref{tab:nu_wind}. Like for the case of the dynamic ejecta, we expect that a more 
detailed treatment would yield shorter times to peak and higher temperatures and luminosities, 
due to geometric effects related to the photosphere that are not captured in the used
semianalytic model. Given the importance of EM signatures of compact 
binary mergers for GW detections such transients from neutrino-driven winds
deserve more detailed studies in future work.

\begin{table}
  \caption{Summary of the results from the neutrino-driven wind model. The mass in the wind is parametrized from $10^{-4}$ to $10^{-2} 
\msun$.}\label{tab:nu_wind}
\vspace*{0.3cm}
\begin{tabular}{@{}llclcl@{}}
\hline
   Run         &  $m_{\rm wind}$ & $t_{\rm peak}$ & $L_{\rm peak}$ & $T_{\rm eff}$ &   \\
   $(m_1-m_2)$ &  $(\msun)$      & (d)            &    (erg/s)     & (K)    &   \\
\hline 
               & $10^{-2}$ & 0.36 & $1.7\times 10^{41}$ & 7700 \\
 A $(1.4-1.4)$ & $10^{-3}$ & 0.25 & $1.1\times 10^{41}$ & 8700 \\
               & $10^{-4}$ & 0.13 & $3.7\times 10^{40}$ & 9900 \\
\hline
               & $10^{-2}$ & 0.46 & $1.8\times 10^{41}$ & 7000 \\
 B $(1.3-1.4)$ & $10^{-3}$ & 0.27 & $1.0\times 10^{41}$ & 8500 \\
               & $10^{-4}$ & 0.13 & $3.5\times 10^{40}$ & 9800 \\
\hline
               & $10^{-2}$ & 0.27 & $1.1\times 10^{40}$ & 7800 \\
 C $(1.6-1.2)$ & $10^{-3}$ & 0.22 & $7.8\times 10^{40}$ & 8400 \\
               & $10^{-4}$ & 0.11 & $3.1\times 10^{40}$ & 10000\\
\hline                                  
               & $10^{-2}$ & 0.07 & $1.7\times 10^{40}$ & 9500 \\
 D $(1.8-1.2)$ & $10^{-3}$ & 0.06 & $1.5\times 10^{40}$ & 9800 \\
               & $10^{-4}$ & 0.05 & $1.2\times 10^{40}$ & 10700\\
\hline
\end{tabular}
\end{table}

\begin{figure*}
  \begin{center}
    \includegraphics[width=0.45\textwidth]{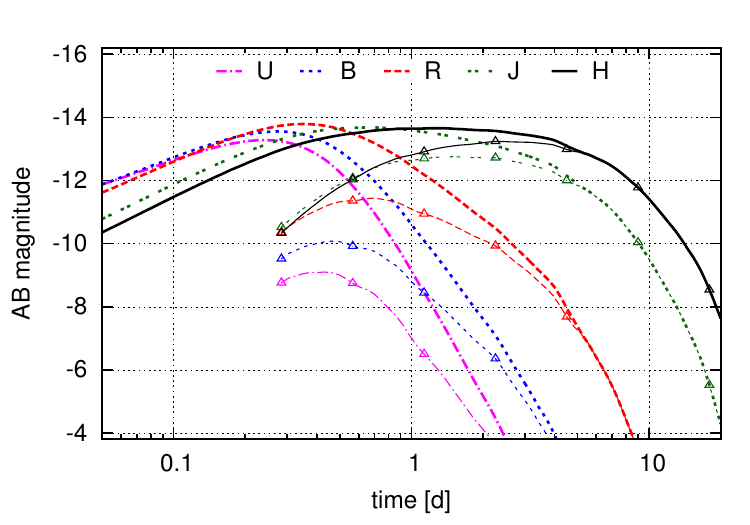}
    \includegraphics[width=0.45\textwidth]{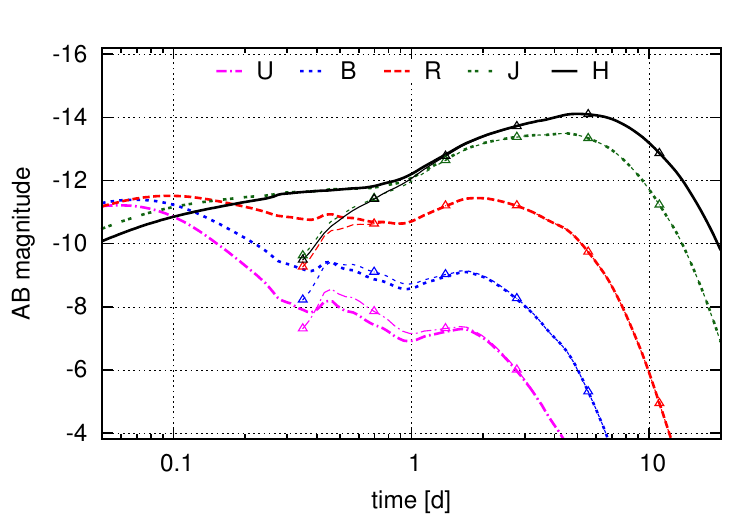}
    \caption{Broad-band light curves from the total ejected material (dynamic ejecta
    plus $\nu$-driven wind; thick lines)  and the dynamically ejected matter alone 
    (symbols). 
    The mass of the $\nu$-driven wind material is set to $10^{-3}\msun$.
    The left-hand panel shows the 1.4--1.3 $\msun$ 
    case, and the right one refers to the 1.8--1.2 $\msun$ case. As a word of caution we 
    stress that the broad-band light curves have been computed assuming a blackbody 
    spectrum corresponding to the effective temperature.
    } 
    \label{fig:broadband-lightcurves}
  \end{center}
\end{figure*}

\begin{figure}
  \begin{center}
    \includegraphics[width=0.45\textwidth]{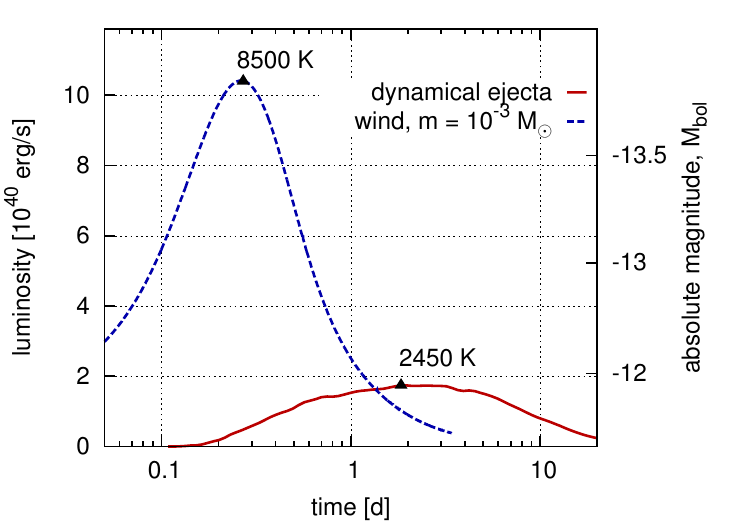}
    \caption{Combined light curves for the 1.3 - 1.4 $\msun$ configuration from the dynamically
    ejected material and the neutrino-driven wind. 
    }
    \label{fig:combined-lightcurves}
  \end{center}
\end{figure}

\section{Detectability}
\label{sec:detect}

Before turning to examine the detectability of the calculated light
curves we note that our IR light curves from the dynamical ejecta, are
dimmer than those of \cite{kasen13a} and of \cite{tanaka13a}. As mentioned
in \S \ref{sec:nucleo} both groups have used radioactive heating rates that
on few days are larger by a factor of 3-4 than our radioactive heating rate
(see fig.~5 of Paper I). We suspect that this is mainly due to the
use of fixed, too large values for the initial electron fraction. Once we correct 
for the effect from the energy generation, our results are in reasonable 
agreement with \cite{barnes13a}. It is worth stressing once more, though, that they
use a more elaborate radiative transfer model (but based on a less sophisticated
model for the matter distribution). In particular, our conclusions are based on
effective temperatures and therefore there is still room for an enhancement 
of the IR part of the spectrum that might enhance the emission in this part of the spectrum.
Overall, our estimates 
of the IR emission from the dynamical ejecta are two magnitudes dimmer than
those of \cite{kasen13a} and of \cite{tanaka13a}, which would make the detectability of
these signals an extremely challenging endeavour.

The most important question concerning these transients is whether they can be 
used to identify ns mergers. Or,  put differently, are the macronova signals detectable?  
When addressing this question two kinds of search modes need to be considered:  
a follow-up search after a GW trigger or a detection in a blind search. In the first case,
the GW trigger provides  an error box of  $\sim100$ deg${^2}$. Given the importance 
of the discovery, one can expect that  significant resources would be allocated for 
this task. Blind transient searches are routine now [e.g. Palomar Transient Factory (PTF) 
and Panoramic Survey Telescope \& Rapid Response System (Pan-STARRS)]  at 
modest magnitudes.  Deeper or high cadence optical searches will  continue  at 
accelerated pace with various telescopes [e.g. Zwicky Transient Factory (ZTF)] and will culminate with the 
operation of the  LSST \citep{LSST09}. SASIR \citep{SASIR} is the only planned 
transit search in the IR. We turn now to examine the detectability of macronovae 
in these two modes. 

We begin by a quick summary of the expected signals. Table \ref{tab: emissions} 
describes the expected signals for the 1.4--1.3 and 1.8--1.2~$\msun$ cases.
These are the weakest and strongest signals and other configurations or signals 
from preferred directions can be stronger by about a factor of 2 (corresponding to 
about one magnitude). 
Fig.~\ref{fig:broadband-lightcurves} shows the broad-band light curves for
these two cases, calculated assuming that radiation is emitted with a 
blackbody spectrum  filtered with the standard filter functions for the
{\it UBVRI} \citep{bessell12} and {\it JHK} \citep{cohen03} bands.
Before turning to these results we stress once more that 
temperatures  estimated here are just effective temperatures and should be taken 
as such. It may very well be that the real signal would be non-thermal and heavily 
dominated by unknown  details of the opacity, radiation transfer and photon emission 
processes. This can modify significantly the detectability estimates presented here. 

The first source is the main IR signal from the dynamical ejecta. It peaks around 
3--6~d after the ns$^2$ merger, with a peak absolute magnitude of (-12.8$,\dots,$ -13.9). 
 The temperature  at this time is about 2300--1800~K  and  most of the emissions will 
 be radiated in the {\it J}--{\it H} band ($ 1.26 \pm 0.1$, $ 1.65 \pm 0.13 \mu m$).  Initially, the temperatures 
 of this component are higher than 4000~K for about half a day, see Fig.~\ref{fig:temperature}. This 
 early signal from the dynamically ejected outflow would be in the visible,   provided that  no
line blanketing occurs. It is relatively weak (-11.6$,\dots,$ -10.8), thus it is of no surprise that 
it would be hard to detect. 
Another source is the emission from the $\nu$-driven winds (see
Fig.~\ref{fig:combined-lightcurves}).
The  emission from the $\nu$-driven winds has an 
absolute magnitude of  (-11.2$,\dots,$ -13.7), depending on the poorly known total mass ejected. Although 
it is short lived (0.4--0.8~d), its high temperature (8000~K) and high luminosity make it a more 
prominent signal and its detection much easier. We consider two possible $\nu$ wind 
cases,  $0.01 \msun$ and of $10^{-3} \msun$, respectively. These signals seem to be
 the strongest among the three considered, but they are about one and a half magnitudes 
 weaker than the previous estimates for macronovae signals that were based on a 
 lower opacity of $\kappa = 0.1 $cm$^2$/g 
\cite[e.g.][]{metzger10b,goriely11a,roberts11,metzger12a,bauswein13a,piran13a}.

\begin{table*}
   \caption{Parameters of the transients, associated with dynamical ejecta and
   $\nu$-driven winds, for two extreme cases: 1.3--1.4~$\msun$ and
   1.8--1.2~$\msun$. $M_{\rm p}$ is the absolute AB magnitude in the relevant band of
   the detector, $m_{\rm p}$ is the magnitude at 300~Mpc, $t_{\rm p}$ is the time since
   merger, and $T_{\rm p}$ is the estimated effective temperature at this time. 
   The last column lists horizon distances for the 5-$\sigma$ detection with
   the exposure of 100~s.
   } 
  \begin{tabular}{|l|c|c|c|c|c|}
  \hline
  Source                             & $M_{\rm p}$(AB) & $t_{\rm p}$  & $T_{\rm p}$ & $m_{\rm p}$(AB) & Horizon (Mpc)\\ 
  ($m_1-m_2$)                        & / Band             & (d)             & (K)           & (@ 300 Mpc)        &  / Inst (@ 100 s).\\ 
  \hline\hline
  Dynamical ejecta:                  &             &        &         &             &               \\ 
  \hline
    1.4--1.3 $\msun$, Main phase      & -12.8 /$J$  & 2      & 2300    & 24.6        & 72 / VISTA    \\ 
                                     &             &        &         &             & 144 / SASIR   \\ 
    1.8--1.2 $\msun$, Main phase      & -13.9 /$H$  & 6      & 1860    & 23.4        & 95 / VISTA    \\ 
                                     &             &        &         &             & 208 / SASIR   \\ 
    1.4--1.3 $\msun$, Initial phase   & -11.6 /$i$  & 0.6    & 4000    & 25.8        & 174 / LSST    \\ 
                                     &             &        &         &             & and Pan-STARRS \\ 
    1.8--1.2 $\msun$, Initial phase   & -10.8 /$i$  & 0.7    & 4000    & 26.6        & 120 / LSST    \\ 
                                     &             &        &         &             & and Pan-STARRS \\ 
  \hline
  $\nu$-driven winds: \\
  \hline
  1.4--1.3 $\msun$, $m_{\rm ej}=10^{-2}\msun$ &-14.3 /$r$& 0.8    & 7676 & 23.0        & 847 / LSST    \\
                                             &-14.6 /$B$&        &      & 23.3        & 650 / HSC$^\dag$  \\ 
  1.8--1.2 $\msun$, $m_{\rm ej}=10^{-2}\msun$ &-11.5 /$r$& $<$0.3 & 9480 & 25.8        & 231 / LSST    \\
                                             &-11.5 /$B$&        &      & 25.8        & 201 / HSC$^\dag$  \\ 
  1.4--1.3 $\msun$, $m_{\rm ej}=10^{-3}\msun$ &-13.7 /$r$& 0.4    & 8740 & 23.7        & 620 / LSST    \\ 
                                             &-13.5 /$B$&        &      & 23.8        & 516 / HSC$^\dag$  \\ 
  1.8--1.2 $\msun$, $m_{\rm ej}=10^{-3}\msun$ &-11.3 /$r$& $<$0.3 & 9800 & 26.0        & 213 / LSST    \\
                                             &-11.4 /$B$&        &      & 26.0        & 188 / HSC$^\dag$  \\ 
  \hline\hline
  \end{tabular}
\newline
$^\dag$ The Hypersuprime camera (HSC) on the Subaru telescope.
\label{tab: emissions}
\end{table*}

Table \ref{tab: Detectors} summarizes the characteristics of a few relevant detectors. 
VISTA \citep{VISTA06} is the largest wide field IR telescope operating today. With its 4~m
mirror it can reach 21.5~mag in a 100~s  exposure. In the optical, Subaru is the 
largest wide-field telescope available; with Hypersuprime camera it would easily reach 25th 
mag over a field of view (FoV) of $\approx 1.5~$deg$^2$, and it is most suitable for a follow-up
survey. Pan-STARRS \citep{kaiser02} is an active transient survey with a limiting magnitude 
of 24.6 and a FoV of 3~deg$^2$. Zwicky Transient Factory (ZTF)
that will replace the PTF is a planned shallow survey ($\approx 21$ mag) 
with a huge FoV (of the order of 30~deg$^2$). The ultimate LSST \citep{LSST09} with a FoV
of 9.6 deg$^2$, will reach a typical limiting magnitude of 24 . 

\begin{table}
\caption{A few telescopes that can search for these transients. The magnitude
  given is for a 5 $\sigma$ detection and for an exposure time of  $t_{\rm
  exp}=100$~s. One can go deeper with the detection magnitude satisfying
  roughly $m_{\rm lim} (t) \approx m_{\rm lim}(t_0)- 2.5 \log[(t/t_0)^{1/2}]$.
} 
\begin{tabular}{|l||c|c|c|c|}
  \hline
  Telescope           & Band & $M_l$(AB)               & FoV     \\
                      &      & For $t_{\rm exp}=100$~s & (deg$^2$) \\ \hline\hline 
  VISTA               & $J$  & 21.5                    & 2.2     \\ \hline 
  SASIR               & $J$  & 23                      & 0.2-1.7 \\ \hline 
  Subaru (Hyper-      & $B$  & 25                      & 1.5     \\
  suprime camera)$^*$ &      &                         &         \\ \hline 
  LSST                & $r$  & 25.3\vspace{1mm}        & 9.6     \\
                      & $i$  & 24.6                    &         \\ \hline 
  ZTP                 & -    & 21.6                    & 30      \\ \hline 
  Pan-STARRS          & $i$  & 24.6                    & 3       \\ \hline 
  \end{tabular} \newline 
  $^*$ We use the Hypersuprime camera FoV with sensitivity estimated from the suprimecam parameters. The 
  Hypersuprime camera sensitivity might be slightly better. 
 \newline
  -- ZFT filters are undecided as yet. 
 \label{tab: Detectors}
  \end{table}

Turning now to detectability, we find that the observations are challenging. 
A follow-up  detection requires covering of a 100~deg$^2$ region at an $\approx 24-25$ mag in the IR for the  
main signal from the dynamic ejecta\footnote{A more sophisticated search can 
cover only a fraction of this area focusing on regions with nearby galaxies~\citep[e.g.][]{Metzger+13}. We do not discuss the details of this 
strategy here.}, at $\approx 25-26$ mag for the initial optical signal from the dynamic
and at $23-24$ mag for the $\nu$-wind signal.  From those, the first main signal in the IR is impossible to detect. 
The optical signals are feasible, but require some effort. One can cover, for example, the whole 100 deg$^2$ 
using the Subaru Hypersuprime camera at 25th B mag within an hour. Thus one can obtain even two coverings of the 
region looking for rapid variability. Alternatively one can cover this region by multiple rapid observations of 
ZFT reaching within one night a 24th magnitude.  Once such a source is found, a follow-up at 
$\approx 24-25$th magnitude at IR  is extremely challenging from Earth (this might be  possible with a several 
hours exposure on VISTA, but doable from space). Such a combined strategy seems to be optimal.
The flat light curve lasting a few days could allow some time for this IR observation. 

With these magnitudes, detection in blind surveys seems hopeless. Even with the LSST, one could expect 
only a few detections per year (assuming an event rate of $300$/yr/Gpc$^3$).  These will be single-event
detections, and unless rapidly identified (from this single observation) and followed up by further optical and 
IR observations, it will be impossible to confirm. Without much more detailed information on the spectral and 
to some extent temporal characteristics of these transients, it would be an impossible task to identify suitable 
candidates within the huge number of transients expected in the LSST. 

\section{Summary and discussion}
\label{sec:conclusion}

In this set of two papers we have investigated the long-term evolution of ns merger
remnants. In the companion paper \citep{rosswog13c}, Paper~I, we have followed the long-term 
evolution of the ejected matter for up to 100 years after the merger. We paid particular 
attention to the role that the heating from radioactively decaying r-process nuclei plays. 
We found that the matter that is dynamically ejected (by tidal and/or hydrodynamic interaction) 
undergoes a `strong' r-process producing a robust pattern of nuclei with $A>130$. Although 
the nuclear energy input does alter dynamics and morphology it does 
not erase the memory of the initial binary mass ratio. We have further 
calculated in a simple model the nucleosynthesis for the expected neutrino-driven wind.
We find that also this wind undergoes an r-process, though a `weak' one which produces
abundance distributions in the range $50 \la A \la 130$ that vary substantially between different
merger  cases. This is because the neutrino luminosities in the different cases yield different $Y_e$ values
within the wind ($0.28 \la Y_e \la 0.40$) and the nucleosynthesis is in this parameter range 
rather sensitive to the exact value of the electron fraction. 

Here we have explored in some detail the properties of radioactively powered
transients from both dynamic ejecta and neutrino-driven winds.
All calculations of the macronova light curves for dynamic ejecta that have been
performed so far have assumed that the expansion is homologous. In Paper~I, this assumption is
dropped and we have explicitly followed the long-term evolution of the dynamical ejecta
including radioactive heating. 
The ejecta composition within the dynamic ejecta is unique and very different from any
type of supernova. Supernovae produce elements up to the iron group near $Z =
26$, but the dynamic ejecta of ns mergers consist entirely of
r-process elements up to the third peak near $Z \approx 90$ and
should thus produce unique features in the EM signal. As recently
argued based on atomic structure calculations by \cite{kasen13a}, this 
material has orders of magnitudes larger opacities than the iron-group-like
values that have been used in previous work. The recent work by \cite{kasen13a} and
\cite{tanaka13a} favours values around 10 cm$^2$/g and this is the value that we have
adopted in the current study. Overall, these higher opacities lead to later, weaker and
redder emission in comparison to earlier macronova models,
but calculations based on the true matter distribution are substantially
brighter, peak earlier and at higher effective temperatures than the
corresponding models of \cite{li98}. This is because the latter assume a
uniform matter distribution and therefore cannot capture geometric and dynamic
properties of the radiating volume. We identify this volume as the location between 
a diffusion surface (where the diffusion time equals the dynamical time) and
the photospheric surface with $\tau=2/3$. In addition, there is a radiation component 
from the optically thin region with $\tau < 2/3$. These photons are produced by
radioactive decays and escape with a spectrum that is unknown to us.

Using the nucleosynthesis results from our simple neutrino-driven wind model we
also estimate the light curves of the corresponding EM transients that arise 
from this component of the outflow. According to our estimates these winds 
should have electron fractions substantially below 0.5  and therefore they 
do not produce any $^{56}$Ni. Still they produce radioactive long-lived enough isotopes that lead to 
radioactive transients.  These isotopes have a substantially lower opacity, as compared 
with the opacity of the dynamically ejected outflow. The resulting transients peak, therefore,  
at very early times ($< 0.5$~d) and are substantially brighter and hotter than the 
transients produced by the dynamic ejecta (see Tab.~\ref{tab:nu_wind}).  Their $\nu$-wind 
transients are relatively bright  ($\sim 10^{41}$ erg/s) with effective temperatures from 
5000 to 10000  K and they proceed the redder and longer lived dynamical  ejecta 
transients. It is worth stressing, however, that the amount of mass in these winds
is not well known and we have parametrized it here with values ranging from $10^{-4}$ to $10^{-2} \; \msun$.
The actual signal depends, of course, on this mass. 

The prospects for unambiguous detection, however, seem  not very optimistic. 
 For the numbers we consider most plausible, the detection prospects
for the dynamic ejecta transients seem dim, even the more promising signals
from the neutrino-driven winds seem one and a half magnitudes lower than previous dynamic
ejecta predictions based on the iron-group-like opacities.
Both signals are too dim to be identified in blind searches. Such searches are carried out in the optical 
and as such the $\nu$-driven wind component is the only one having a chance of being detected. 
However, with reasonable estimates of the merger event rate even the mammoth LSST will detect only a few such
transients per year and with a single detection these are most likely to be lost within millions of other transients discovered. 
The situation is somewhat better with a GW trigger. Such a trigger will provide an $\sim 100 $ deg$^2$ error box. 
Even with this information, a search for the IR signal from the dynamical ejecta is impossible with 
current technology. {\it WFIRST} \citep{WFIRST} is the only mission that could possibly perform such 
a search, but it is far from clear that it will ever be launched. Under optimistic conditions the 
transient from  a significant ($10^{-2} \; \msun$) $\nu$-driven wind can then be possibly detected, for example using 
the Hypersuprime camera on Subaru, in a dedicated follow-up search that will cover the whole 
error box with a sufficient depth.  With some luck this UV/optical  can be detected even using  a 
continuous exposure of a smaller telescope (like the ZFT) that has  a very wide FoV.  
Once such a transient is detected and localized an IR follow-up is possible. This will require 
quick allocation of a significant amount of time on the best IR facilities available,  but the 
stakes are high. A combined detection of both signals would provide a `smoking gun' signature 
confirming the nature of the transient as an ns merger-driven macronova. 
 
\cite{tanvir13}
and \cite{berger13} discovered recently, using the {\it Hubble Space Telescope}, an nIR transient
at the location of the short burst GRB~130603B. This transient had an apparent magnitude of 
$25.3 \pm 0.3$ corresponding to $M_{J,{\rm AB}}\approx-15.35$ 
approximately seven days after the burst in the progenitor rest frame.
Both groups suggest that, while not fully conclusive, 
this could have been a macronova signature of a merger event. 
The observed magnitude is consistent with the models of
\cite{barnes13a,tanaka13a} and \cite{hotokezaka13b}.
However, as mentioned earlier (\S \ref{sec:nucleo}), we believe that the nuclear heating rate used in these 
models and hence the observed flux at the crucial time is overestimated 
by a factor of $3-4$ and as our average estimates are 2 mag below this observed transient. 
On the other hand, as the GRB was observed, it is natural to assume
that the system had a  favourable orientation (the `top' view). This would
enhance the luminosity by a factor of 2 relative to the average one. 
Our estimated signal would still fall short by about 1~mag compared with the 
observations\footnote{
  Additionally, the material from dissolution of the accretion disc
  (not considered here) will have higher $Y_e\sim0.2$ and correspondingly
  higher nuclear energy release at late times (see fig.6 of Paper~I). If the
  amount of mass ejected in this channel is comparable to the dynamically
  ejected mass, this can affect our detectability estimates.
}. 
While this difference  is within what we consider  
a reasonable uncertainty of our estimates, significantly more massive 
dynamical ejecta, as would have been the case in a dynamic nsbh collision
\citep{lee10a,rosswog13a}, some nsbh mergers 
\citep{rosswog05a,deaton13,foucart13,hotokezaka13b,kyutoku13,lovelace13,tanaka13b} 
or a higher $Y_e$ (which could arise in a disruption of a particularly small 
ns with a large crust) could possibly bridge this gap.

\section*{Acknowledgements}
We would like to thank C. Winteler for providing his nucleosynthesis network
code and for his continued support and for helpful comments. 
This work has also benefited from the stimulating discussions at the MICRA workshop in 2013.
We gratefully acknowledge helpful discussions with Mattias Ergon, Claes Fransson, Ariel Goobar,
Ehud Nakar, Dovi Poznanski, Jesper Sollerman and  Ivan Zalamea.
DG and TP were supported by an ERC advanced grant (GRBs) and by the  I-CORE 
Programme of the Planning and Budgeting Committee and The Israel Science Foundation (grant no. 1829/12).
SR and OK were supported by DFG grant RO-3399,  AOBJ-584282 and by the Swedish 
Research Council (VR) under grant 621-2012-4870.  SR has  been supported
by Compstar. The simulations of this paper were performed on the facilities of the
H\"ochstleistungsrechenzentrum Nord (HLRN).

\hyphenation{Post-Script Sprin-ger}
%\bibliographystyle{mn2e_eprint}
%\bibliography{paper_II}

\begin{thebibliography}{}

\bibitem[\protect\citeauthoryear{{Abadie}, {Abbott}, {Abbott}, {Abernathy},
  {Accadia}, {Acernese}, {Adams}, {Adhikari}, {Ajith}, {Allen} \& et
  al.}{{Abadie} et~al.}{2010}]{abadie10}
{Abadie} J. et al., 2010, CQG, 27, 173001, 
  \adsurl{http://adsabs.harvard.edu/abs/2010CQGra..27q3001A}

\bibitem[\protect\citeauthoryear{{Accadia}, {Acernese}, {Antonucci}, {Astone},
  {Ballardin}, {Barone}, {Barsuglia}, {Basti}, {Bauer}, {Bebronne}, {Beker},
  {Belletoile} A.~{Ward}, {Was}, {Yvert} \& {Zendri}}{{Accadia}
  et~al.}{2011}]{accadia11}
{Accadia} T. et al.,  2011, CQG, 28, 114002,
  \adsurl{http://adsabs.harvard.edu/abs/2011CQGra..28k4002A}

\bibitem[\protect\citeauthoryear{{Aloy}, {Janka} \& {M{\"u}ller}}{{Aloy}
  et~al.}{2005}]{aloy05}
{Aloy} M.~A.,  {Janka} H.-T.,    {M{\"u}ller} E.,  2005, A\&A, 436, 273,
  \adsurl{http://adsabs.harvard.edu/abs/2005A\&A...436..273A}

\bibitem[\protect\citeauthoryear{{Arun}, {Babak}, {Berti}, {Cornish}, {Cutler},
  {Gair}, {Hughes}, {Iyer}, {Lang}, {Mandel}, {Porter}, {Sathyaprakash},
  {Sinha}, {Sintes}, {Trias}, {Van Den Broeck} \& {Volonteri}}{{Arun}
  et~al.}{2009}]{arun09}
{Arun} K.~G. et al.,  2009, CQG, 26, 094027, 
  \adsurl{http://adsabs.harvard.edu/abs/2009CQGra..26i4027A}

\bibitem[\protect\citeauthoryear{{Barnes} \& {Kasen}}{{Barnes} \&
  {Kasen}}{2013}]{barnes13a}
{Barnes} J.,  {Kasen} D.,  2013, ApJ, 775, 18, 
  \adsurl{http://adsabs.harvard.edu/abs/2013ApJ...775...18B}

\bibitem[\protect\citeauthoryear{{Bartos}, {Brady} \& {M{\'a}rka}}{{Bartos}
  et~al.}{2013}]{bartos13}
{Bartos} I.,  {Brady} P.,    {M{\'a}rka} S.,  2013, CQG, 30, 123001, 
  \adsurl{http://adsabs.harvard.edu/abs/2013CQGra..30l3001B}

\bibitem[\protect\citeauthoryear{{Bauswein}, {Goriely} \& {Janka}}{{Bauswein}
  et~al.}{2013}]{bauswein13a}
{Bauswein} A.,  {Goriely} S.,    {Janka} H.-T.,  2013, ApJ, 773, 78,
  \adsurl{http://adsabs.harvard.edu/abs/2013ApJ...773...78B}

\bibitem[\protect\citeauthoryear{{Beloborodov}}{{Beloborodov}}{2008}]{beloboro%
dov08}
{Beloborodov} A.~M.,  2008, in {M.~Axelsson} ed., AIPConf. Ser. Vol.~1054,
  {Hyper-accreting Black Holes}.
pp 51--70, \adsurl{http://adsabs.harvard.edu/abs/2008AIPC.1054...51B}

\bibitem[\protect\citeauthoryear{{Berger}}{{Berger}}{2009}]{berger09}
{Berger} E.,  2009, ApJ, 690, 231, 
  \adsurl{http://adsabs.harvard.edu/abs/2009ApJ...690..231B}

\bibitem[\protect\citeauthoryear{{Berger}}{{Berger}}{2010}]{berger10}
{Berger} E.,  2010, ApJ, 722, 1946, 
  \adsurl{http://adsabs.harvard.edu/abs/2010ApJ...722.1946B}

\bibitem[\protect\citeauthoryear{{Berger}, {Fong} \& {Chornock}}{{Berger}
  et~al.}{2013}]{berger13}
{Berger} E.,  {Fong} W.,    {Chornock} R.,  2013, ApJ, 774, L23,
  \adsurl{http://adsabs.harvard.edu/abs/2013ApJ...774L..23B}
   
\bibitem[\protect\citeauthoryear{{Bessell} \& {Murphy}}{{Bessell} \&
  {Murphy}}{2012}]{bessell12}
{Bessell} M.,  {Murphy} S.,  2012, PASP, 124, 140, 
  \adsurl{http://adsabs.harvard.edu/abs/2012PASP..124..140B}

\bibitem[\protect\citeauthoryear{{Bloom} et al.}{{Bloom} \&
  et~al.}{2009}]{bloom09b}
{Bloom} J.~S. et al. 2009, \eprint{0902.1527}

\bibitem[\protect\citeauthoryear{{Bloom}, {Prochaska}, {Lee}, {Jes{\'u}s
  Gonz{\'a}lez}, {Ram{\'{\i}}rez-Ruiz}, {Bolte}, {Franco}, {Guichard},
  {Carrami{\~n}ana}, {Strittmatter}, {Avila-Reese} \& other authors}{{Bloom}
  et~al.}{2009}]{SASIR}
{Bloom} J.~S. et al. 2009, \eprint{0905.1965}

\bibitem[\protect\citeauthoryear{{Caballero}, {McLaughlin} \&
  {Surman}}{{Caballero} et~al.}{2012}]{caballero12}
{Caballero} O.~L.,  {McLaughlin} G.~C.,    {Surman} R.,  2012, ApJ, 745, 170,
  \adsurl{http://adsabs.harvard.edu/abs/2012ApJ...745..170C}

\bibitem[\protect\citeauthoryear{{Cohen}, {Wheaton} \& {Megeath}}{{Cohen}
  et~al.}{2003}]{cohen03}
{Cohen} M.,  {Wheaton} W.~A.,    {Megeath} S.~T.,  2003, AJ, 126, 1090,
  \adsurl{http://adsabs.harvard.edu/abs/2003AJ....126.1090C}

\bibitem[\protect\citeauthoryear{{Coward}, {Howell}, {Piran}, {Stratta},
  {Branchesi}, {Bromberg}, {Gendre}, {Burman} \& {Guetta}}{{Coward}
  et~al.}{2012}]{coward12}
{Coward} D.~M. et al. 2012, MNRAS,
  425, 2668, \adsurl{http://adsabs.harvard.edu/abs/2012MNRAS.425.2668C}

\bibitem[\protect\citeauthoryear{{Dalal}, {Holz}, {Hughes} \& {Jain}}{{Dalal}
  et~al.}{2006}]{dalal06}
{Dalal} N.,  {Holz} D.~E.,  {Hughes} S.~A.,    {Jain} B.,  2006, Phys. Rev. D,
  74, 063006, \adsurl{http://adsabs.harvard.edu/abs/2006PhRvD..74f3006D}

\bibitem[\protect\citeauthoryear{{Deaton}, {Duez}, {Foucart}, {O'Connor},
  {Ott}, {Kidder}, {Muhlberger}, {Scheel} \& {Szilagyi}}{{Deaton}
  et~al.}{2013}]{deaton13}
{Deaton} M.~B. et al. 2013, ApJ, 776, 47, 
  \adsurl{http://adsabs.harvard.edu/abs/2013ApJ...776...47D}

\bibitem[\protect\citeauthoryear{{Dessart}, {Ott}, {Burrows}, {Rosswog} \&
  {Livne}}{{Dessart} et~al.}{2009}]{dessart09}
{Dessart} L.,  {Ott} C.~D.,  {Burrows} A.,  {Rosswog} S.,    {Livne} E.,  2009,
  ApJ, 690, 1681, \adsurl{http://adsabs.harvard.edu/abs/2009ApJ...690.1681D}

\bibitem[\protect\citeauthoryear{{Eichler}, {Livio}, {Piran} \&
  {Schramm}}{{Eichler} et~al.}{1989}]{eichler89}
{Eichler} D.,  {Livio} M.,  {Piran} T.,    {Schramm} D.~N.,  1989, Nature, 340,
  126, \adsurl{http://adsabs.harvard.edu/abs/1989Natur.340..126E}

\bibitem[\protect\citeauthoryear{{Emerson}, {McPherson} \&
  {Sutherland}}{{Emerson} et~al.}{2006}]{VISTA06}
{Emerson} J.,  {McPherson} A.,    {Sutherland} W.,  2006, The Messenger, 126,
  41, \adsurl{http://adsabs.harvard.edu/abs/2006Msngr.126...41E}

\bibitem[\protect\citeauthoryear{{Fern{\'a}ndez} \& {Metzger}}{{Fern{\'a}ndez}
  \& {Metzger}}{2013}]{fernandez13}
{Fern{\'a}ndez} R.,  {Metzger} B.~D.,  2013, MNRAS, 435, 502,
  \adsurl{http://adsabs.harvard.edu/abs/2013MNRAS.435..502F}

\bibitem[\protect\citeauthoryear{{Fong}, {Berger} \& {Fox}}{{Fong}
  et~al.}{2010}]{fong10}
{Fong} W.,  {Berger} E.,    {Fox} D.~B.,  2010, ApJ, 708, 9,
  \adsurl{http://adsabs.harvard.edu/abs/2010ApJ...708....9F}

\bibitem[\protect\citeauthoryear{{Fong}, {Berger}, {Margutti}, {Zauderer},
  {Troja}, {Czekala}, {Chornock}, {Gehrels}, {Sakamoto}, {Fox} \&
  {Podsiadlowski}}{{Fong} et~al.}{2012}]{fong12}
{Fong} W. et al.  2012, ApJ, 756, 189, 
  \adsurl{http://adsabs.harvard.edu/abs/2012ApJ...756..189F}

\bibitem[\protect\citeauthoryear{{Fong}, {Berger}, {Servillat}, {Anglada} \&
  more authors}{{Fong} et~al.}{2013}]{fong13}
{Fong} W. et al. 2013, ApJ, 769, 56, 
  \adsurl{http://adsabs.harvard.edu/abs/2013ApJ...769...56F}

\bibitem[\protect\citeauthoryear{{Foucart}, {Deaton}, {Duez}, {Kidder},
  {MacDonald}, {Ott}, {Pfeiffer}, {Scheel}, {Szilagyi} \&
  {Teukolsky}}{{Foucart} et~al.}{2013}]{foucart13}
{Foucart} F. et al.  2013, Phys. Rev. D, 87, 084006, 
  \adsurl{http://adsabs.harvard.edu/abs/2013PhRvD..87h4006F}

\bibitem[\protect\citeauthoryear{{Freiburghaus}, {Rosswog} \&
  {Thielemann}}{{Freiburghaus} et~al.}{1999}]{freiburghaus99b}
{Freiburghaus} C.,  {Rosswog} S.,    {Thielemann} F.-K.,  1999, ApJ, 525, L121,
  \adsurl{http://adsabs.harvard.edu/abs/1999ApJ...525L.121F}

\bibitem[\protect\citeauthoryear{{Goriely}, {Bauswein} \& {Janka}}{{Goriely}
  et~al.}{2011}]{goriely11a}
{Goriely} S.,  {Bauswein} A.,    {Janka} H.-T.,  2011, ApJL, 738, L32,
  \adsurl{http://adsabs.harvard.edu/abs/2011ApJ...738L..32G}

\bibitem[\protect\citeauthoryear{{Guetta} \& {Piran}}{{Guetta} \&
  {Piran}}{2006}]{guetta06}
{Guetta} D.,  {Piran} T.,  2006, A\&A, 453, 823,
  \adsurl{http://adsabs.harvard.edu/abs/2006A\&A...453..823G}

\bibitem[\protect\citeauthoryear{{Guetta} \& {Stella}}{{Guetta} \&
  {Stella}}{2009}]{guetta09}
{Guetta} D.,  {Stella} L.,  2009, A\&A, 498, 329, 
  \adsurl{http://adsabs.harvard.edu/abs/2009A\&A...498..329G}

\bibitem[\protect\citeauthoryear{{Harry} \& {LIGO Scientific
  Collaboration}}{{Harry} \& {LIGO Scientific Collaboration}}{2010}]{harry10}
{Harry} G.~M. et al. 2010, CQG, 27, 084006,
  \adsurl{http://adsabs.harvard.edu/abs/2010CQGra..27h4006H}

\bibitem[\protect\citeauthoryear{{Hoffman}, {Woosley} \& {Qian}}{{Hoffman}
  et~al.}{1997}]{hoffman97}
{Hoffman} R.~D.,  {Woosley} S.~E.,    {Qian} Y.-Z.,  1997, ApJ, 482, 951,
  \adsurl{http://adsabs.harvard.edu/abs/1997ApJ...482..951H}

\bibitem[\protect\citeauthoryear{{Hotokezaka}, {Kiuchi}, {Kyutoku}, {Okawa},
  {Sekiguchi}, {Shibata} \& {Taniguchi}}{{Hotokezaka}
  et~al.}{2013}]{hotokezaka13a}
{Hotokezaka} K.,  {Kiuchi} K.,  {Kyutoku} K.,  {Okawa} H.,  {Sekiguchi} Y.,
  {Shibata} M.,    {Taniguchi} K.,  2013, Phys. Rev. D, 87, 024001,
  \adsurl{http://adsabs.harvard.edu/abs/2013PhRvD..87b4001H}

\bibitem[\protect\citeauthoryear{{Hotokezaka}, {Kyutoku}, {Tanaka}, {Kiuchi},
  {Sekiguchi}, {Shibata} \& {Wanajo}}{{Hotokezaka}
  et~al.}{2013}]{hotokezaka13b}
{Hotokezaka} K.,  {Kyutoku} K.,  {Tanaka} M.,  {Kiuchi} K.,  {Sekiguchi} Y.,
  {Shibata} M.,    {Wanajo} S.,  2013, ApJ, 778, L16, 
  \adsurl{http://adsabs.harvard.edu/abs/2013ApJ...778L..16H}

\bibitem[\protect\citeauthoryear{{Hughes} \& {Holz}}{{Hughes} \&
  {Holz}}{2003}]{hughes03}
{Hughes} S.~A.,  {Holz} D.~E.,  2003, CQG, 20, 65,
  \adsurl{http://adsabs.harvard.edu/abs/2003CQGra..20S..65H}

\bibitem[\protect\citeauthoryear{{Kaiser}, {Aussel}, {Burke}, {Boesgaard},
  {Chambers}, {Chun}, {Heasley}, {Hodapp}, {Hunt}, {Jedicke}, {Jewitt},
  {Kudritzki} \& other authors}{{Kaiser} et~al.}{2002}]{kaiser02}
{Kaiser} N. et al. 2002, in {J.~A.~Tyson \&
  S.~Wolff} ed., {Proc. SPIE Conf. Ser. 4836}, {Pan-STARRS: A Large Synoptic 
  Survey Telescope Array}.
p 154, \adsurl{http://adsabs.harvard.edu/abs/2002SPIE.4836..154K}

\bibitem[\protect\citeauthoryear{{Kalogera}, {Kim}, {Lorimer}, {Burgay},
  {D'Amico}, {Possenti}, {Manchester}, {Lyne}, {Joshi}, {McLaughlin}, {Kramer},
  {Sarkissian} \& {Camilo}}{{Kalogera} et~al.}{2004b}]{kalogera04b}
{Kalogera} V. et al. 2004b, ApJ, 614, L137, 
  \adsurl{http://adsabs.harvard.edu/abs/2004ApJ...614L.137K}

\bibitem[\protect\citeauthoryear{{Kalogera}, {Kim}, {Lorimer}, {Burgay},
  {D'Amico}, {Possenti}, {Manchester}, {Lyne}, {Joshi}, {McLaughlin}, {Kramer},
  {Sarkissian} \& {Camilo}}{{Kalogera} et~al.}{2004a}]{kalogera04a}
{Kalogera} V. et al. 2004a, ApJ, 601, L179,
  \adsurl{http://adsabs.harvard.edu/abs/2004ApJ...601L.179K}

\bibitem[\protect\citeauthoryear{{Kasen}, {Badnell} \& {Barnes}}{{Kasen}
  et~al.}{2013}]{kasen13a}
{Kasen} D.,  {Badnell} N.~R.,    {Barnes} J.,  2013, ApJ, 774, 25,
  \adsurl{http://adsabs.harvard.edu/abs/2013ApJ...774...25K}

\bibitem[\protect\citeauthoryear{{Kasliwal} \& {Nissanke}}{{Kasliwal} \&
  {Nissanke}}{2013}]{kasliwal13}
{Kasliwal} M.~M.,  {Nissanke} S.,  2013, \eprint{1309.1554},

\bibitem[\protect\citeauthoryear{Kelley, Mandel \& Ramirez-Ruiz}{Kelley
  et~al.}{2012}]{kelley12}
Kelley L.~Z.,  Mandel I.,    Ramirez-Ruiz E.,  2012, Physical Review D, 87, 17,
  \adsurl{http://adsabs.harvard.edu/abs/2013PhRvD..87l3004K}

\bibitem[\protect\citeauthoryear{{Kochanek} \& {Piran}}{{Kochanek} \&
  {Piran}}{1993}]{kochanek93}
{Kochanek} C.~S.,  {Piran} T.,  1993, ApJL, 417, L17,
  \adsurl{http://adsabs.harvard.edu/abs/1993ApJ...417L..17K}

\bibitem[\protect\citeauthoryear{{Korobkin}, {Rosswog}, {Arcones} \&
  {Winteler}}{{Korobkin} et~al.}{2012}]{korobkin12a}
{Korobkin} O.,  {Rosswog} S.,  {Arcones} A.,    {Winteler} C.,  2012, MNRAS,
  426, 1940, \adsurl{http://adsabs.harvard.edu/abs/2012MNRAS.426.1940K}

\bibitem[\protect\citeauthoryear{{Kulkarni}}{{Kulkarni}}{2005}]{kulkarni05}
{Kulkarni} S.~R.,  2005, \eprint{astro-ph/0510256},

\bibitem[\protect\citeauthoryear{{Kupka}, {Ryabchikova}, {Piskunov}, {Stempels}
  \& {Weiss}}{{Kupka} et~al.}{2000}]{kupka00}
{Kupka} F.~G.,  {Ryabchikova} T.~A.,  {Piskunov} N.~E.,  {Stempels} H.~C.,
  {Weiss} W.~W.,  2000, Baltic Astronomy, 9, 590,
  \adsurl{http://adsabs.harvard.edu/abs/2000BaltA...9..590K}

\bibitem[\protect\citeauthoryear{{Kyutoku}, {Ioka} \& {Shibata}}{{Kyutoku}
  et~al.}{2013}]{kyutoku13}
{Kyutoku} K.,  {Ioka} K.,    {Shibata} M.,  2013, Phys. Rev. D, 88, 041503,
  \adsurl{http://adsabs.harvard.edu/abs/2013PhRvD..88d1503K}

\bibitem[\protect\citeauthoryear{{Lattimer}, {Mackie}, {Ravenhall} \&
  {Schramm}}{{Lattimer} et~al.}{1977}]{lattimer77}
{Lattimer} J.~M.,  {Mackie} F.,  {Ravenhall} D.~G.,    {Schramm} D.~N.,  1977,
  ApJ, 213, 225, \adsurl{http://adsabs.harvard.edu/abs/1977ApJ...213..225L}

\bibitem[\protect\citeauthoryear{{Lattimer} \& {Schramm}}{{Lattimer} \&
  {Schramm}}{1974}]{lattimer74}
{Lattimer} J.~M.,  {Schramm} D.~N.,  1974, ApJ, 192, L145,
  \adsurl{http://adsabs.harvard.edu/abs/1974ApJ...192L.145L}

\bibitem[\protect\citeauthoryear{{Lee} \& {Ramirez-Ruiz}}{{Lee} \&
  {Ramirez-Ruiz}}{2007}]{lee07}
{Lee} W.~H.,  {Ramirez-Ruiz} E.,  2007, New Journal of Physics, 9, 17,
  \adsurl{http://adsabs.harvard.edu/abs/2007NJPh....9...17L}

\bibitem[\protect\citeauthoryear{{Lee}, {Ramirez-Ruiz} \&
  {L{\'o}pez-C{\'a}mara}}{{Lee} et~al.}{2009}]{lee09}
{Lee} W.~H.,  {Ramirez-Ruiz} E.,    {L{\'o}pez-C{\'a}mara} D.,  2009, ApJL,
  699, L93, \adsurl{http://adsabs.harvard.edu/abs/2009ApJ...699L..93L}

\bibitem[\protect\citeauthoryear{{Lee}, {Ramirez-Ruiz} \& {van de Ven}}{{Lee}
  et~al.}{2010}]{lee10a}
{Lee} W.~H.,  {Ramirez-Ruiz} E.,    {van de Ven} G.,  2010, ApJ, 720, 953,
  \adsurl{http://adsabs.harvard.edu/abs/2010ApJ...720..953L}

\bibitem[\protect\citeauthoryear{{Li} \& {Paczy{\'n}ski}}{{Li} \&
  {Paczy{\'n}ski}}{1998}]{li98}
{Li} L.,  {Paczy{\'n}ski} B.,  1998, ApJL, 507, L59,
  \adsurl{http://adsabs.harvard.edu/abs/1998ApJ...507L..59L}

\bibitem[\protect\citeauthoryear{{Lovelace}, {Duez}, {Foucart}, {Kidder},
  {Pfeiffer}, {Scheel} \& {Szil{\'a}gyi}}{{Lovelace} et~al.}{2013}]{lovelace13}
{Lovelace} G.,  {Duez} M.~D.,  {Foucart} F.,  {Kidder} L.~E.,  {Pfeiffer}
  H.~P.,  {Scheel} M.~A., {Szil{\'a}gyi} B., 2013, CQG, 30, 135004, 
  \adsurl{http://adsabs.harvard.edu/abs/2013CQGra..30m5004L}

\bibitem[\protect\citeauthoryear{{LSST Science Collaborations}, {Abell},
  {Allison}, {Anderson}, {Andrew}, {Angel}, {Armus}, {Arnett}, {Asztalos},
  {Axelrod} et al.}{{LSST Science Collaborations} et~al.}{2009}]{LSST09}
{LSST Science Collaborations} et al., 2009, \eprint{0912.0201},

\bibitem[\protect\citeauthoryear{{Metzger} \& {Berger}}{{Metzger} \&
  {Berger}}{2012}]{metzger12a}
{Metzger} B.~D.,  {Berger} E.,  2012, ApJ, 746, 48, 
  \adsurl{http://adsabs.harvard.edu/abs/2012ApJ...746...48M}

\bibitem[\protect\citeauthoryear{{Metzger}, {Kaplan} \& {Berger}}{{Metzger}
  et~al.}{2013}]{Metzger+13}
{Metzger} B.~D.,  {Kaplan} D.~L.,    {Berger} E.,  2013, ApJ, 764, 149,
  \adsurl{http://adsabs.harvard.edu/abs/2013ApJ...764..149M}

\bibitem[\protect\citeauthoryear{{Metzger}, {Mart{\'{\i}}nez-Pinedo}, {Darbha},
  {Quataert}, {Arcones}, {Kasen}, {Thomas}, {Nugent}, {Panov} \&
  {Zinner}}{{Metzger} et~al.}{2010}]{metzger10b}
{Metzger} B.~D. et al.  2010, MNRAS, 406, 2650, 
  \adsurl{http://adsabs.harvard.edu/abs/2010MNRAS.406.2650M}

\bibitem[\protect\citeauthoryear{{Metzger}, {Piro} \& {Quataert}}{{Metzger}
  et~al.}{2008}]{metzger08}
{Metzger} B.~D.,  {Piro} A.~L.,    {Quataert} E.,  2008, MNRAS, 390, 781,
  \adsurl{http://adsabs.harvard.edu/abs/2008MNRAS.390..781M}

\bibitem[\protect\citeauthoryear{{Metzger}, {Piro} \& {Quataert}}{{Metzger}
  et~al.}{2009}]{metzger09b}
{Metzger} B.~D.,  {Piro} A.~L.,    {Quataert} E.,  2009, MNRAS, 396, 304,
  \adsurl{http://adsabs.harvard.edu/abs/2009MNRAS.396..304M}

\bibitem[\protect\citeauthoryear{{Monaghan}}{{Monaghan}}{2005}]{monaghan05}
{Monaghan} J.~J.,  2005, Reports on Progress in Physics, 68, 1703,
  \adsurl{http://adsabs.harvard.edu/abs/2005RPPh...68.1703M}

\bibitem[\protect\citeauthoryear{{Nakar}}{{Nakar}}{2007}]{nakar07}
{Nakar} E.,  2007, Phys. Rep., 442, 166, 
  \adsurl{http://adsabs.harvard.edu/abs/2007PhR...442..166N}

\bibitem[\protect\citeauthoryear{{Nakar}, {Gal-Yam} \& {Fox}}{{Nakar}
  et~al.}{2006}]{nakar06}
{Nakar} E.,  {Gal-Yam} A.,    {Fox} D.~B.,  2006, ApJ, 650, 281,
  \adsurl{http://adsabs.harvard.edu/abs/2006ApJ...650..281N}

\bibitem[\protect\citeauthoryear{{Nakar} \& {Piran}}{{Nakar} \&
  {Piran}}{2011}]{NP11}
{Nakar} E.,  {Piran} T.,  2011, Nature, 478, 82,
  \adsurl{http://adsabs.harvard.edu/abs/2011Natur.478...82N}

\bibitem[\protect\citeauthoryear{{Narayan}, {Piran} \& {Shemi}}{{Narayan}
  et~al.}{1991}]{narayan91}
{Narayan} R.,  {Piran} T.,    {Shemi} A.,  1991, ApJ, 379, L17,
  \adsurl{http://adsabs.harvard.edu/abs/1991ApJ...379L..17N}

\bibitem[\protect\citeauthoryear{{Nissanke}, {Kasliwal} \&
  {Georgieva}}{{Nissanke} et~al.}{2013}]{nissanke13}
{Nissanke} S.,  {Kasliwal} M.,    {Georgieva} A.,  2013, ApJ, 767, 124,
  \adsurl{http://adsabs.harvard.edu/abs/2013ApJ...767..124N}

\bibitem[\protect\citeauthoryear{{Phinney}}{{Phinney}}{1991}]{phinney91}
{Phinney} E.~S.,  1991, ApJ, 380, L17,
  \adsurl{http://adsabs.harvard.edu/abs/1991ApJ...380L..17P}

\bibitem[\protect\citeauthoryear{{Phinney}}{{Phinney}}{2009}]{phinney09}
{Phinney} E.~S.,  2009, in Astro2010: The Astronomy and Astrophysics Decadal
  Survey, Vol.~2010, {Finding and Using Electromagnetic
  Counterparts of Gravitational Wave Sources}.
p.~235, \adsurl{http://adsabs.harvard.edu/abs/2009astro2010S.235P}

\bibitem[\protect\citeauthoryear{{Piran}}{{Piran}}{2004}]{piran04}
{Piran} T.,  2004, Reviews of Modern Physics, 76, 1143,
  \adsurl{http://adsabs.harvard.edu/abs/2004RvMP...76.1143P}

\bibitem[\protect\citeauthoryear{{Piran}, {Nakar} \& {Rosswog}}{{Piran}
  et~al.}{2013}]{piran13a}
{Piran} T.,  {Nakar} E.,    {Rosswog} S.,  2013, MNRAS, 430, 2121,
  \adsurl{http://adsabs.harvard.edu/abs/2013MNRAS.430.2121P}

\bibitem[\protect\citeauthoryear{{Qian} \& {Woosley}}{{Qian} \&
  {Woosley}}{1996}]{qian96b}
{Qian} Y.-Z.,  {Woosley} S.~E.,  1996, ApJ, 471, 331,
  \adsurl{http://adsabs.harvard.edu/abs/1996ApJ...471..331Q}

\bibitem[\protect\citeauthoryear{{Roberts}, {Kasen}, {Lee} \&
  {Ramirez-Ruiz}}{{Roberts} et~al.}{2011}]{roberts11}
{Roberts} L.~F.,  {Kasen} D.,  {Lee} W.~H.,    {Ramirez-Ruiz} E.,  2011, ApJL,
  736, L21+, \adsurl{http://adsabs.harvard.edu/abs/2011ApJ...736L..21R}

\bibitem[\protect\citeauthoryear{{Rosswog}}{{Rosswog}}{2005}]{rosswog05a}
{Rosswog} S.,  2005, ApJ, 634, 1202,
  \adsurl{http://adsabs.harvard.edu/abs/2005ApJ...634.1202R}

\bibitem[\protect\citeauthoryear{{Rosswog}}{{Rosswog}}{2009}]{rosswog09b}
{Rosswog} S.,  2009, New Astronomy Reviews, 53, 78, 
  \adsurl{http://adsabs.harvard.edu/abs/2009NewAR..53...78R}

\bibitem[\protect\citeauthoryear{{Rosswog}}{{Rosswog}}{2013}]{rosswog13b}
{Rosswog} S.,  2013, Philosophical Transactions A, 1210.6549,
  \adsurl{http://adsabs.harvard.edu/abs/2013RSPTA.37120272R}

\bibitem[\protect\citeauthoryear{{Rosswog} \& {Davies}}{{Rosswog} \&
  {Davies}}{2002}]{rosswog02a}
{Rosswog} S.,  {Davies} M.~B.,  2002, MNRAS, 334, 481,
  \adsurl{http://adsabs.harvard.edu/abs/2002MNRAS.334..481R}

\bibitem[\protect\citeauthoryear{{Rosswog}, {Davies}, {Thielemann} \&
  {Piran}}{{Rosswog} et~al.}{2000}]{rosswog00}
{Rosswog} S.,  {Davies} M.~B.,  {Thielemann} F.-K.,    {Piran} T.,  2000, A\&A,
  360, 171, \adsurl{http://adsabs.harvard.edu/abs/2000A\&A...360..171R}

\bibitem[\protect\citeauthoryear{{Rosswog}, {Korobkin}, {Arcones} \&
  {Thielemann}}{{Rosswog} et~al.}{2013}]{rosswog13c}
{Rosswog} S.,  {Korobkin} O.,  {Arcones} A., {Thielemann} F.-K., {Piran} T., 2013,

\bibitem[\protect\citeauthoryear{{Rosswog} \& {Liebend{\"o}rfer}}{{Rosswog} \&
  {Liebend{\"o}rfer}}{2003}]{rosswog03a}
{Rosswog} S.,  {Liebend{\"o}rfer} M.,  2003, MNRAS, 342, 673,
  \adsurl{http://adsabs.harvard.edu/abs/2003MNRAS.342..673R}

\bibitem[\protect\citeauthoryear{{Rosswog}, {Liebend{\"o}rfer}, {Thielemann},
  {Davies}, {Benz} \& {Piran}}{{Rosswog} et~al.}{1998}]{rosswog98a}
{Rosswog} S.,  {Liebend{\"o}rfer} M.,  {Thielemann} F.-K.,  {Davies} M.,
  {Benz} W.,    {Piran} T.,  1998, in {Mezzacappa} A.,  ed., Proc. 2nd Oak
  Ridge Symp., Atomic and Nuclear Astrophysics, {Mass ejection in neutron star
  mergers}.
p.~729, \adsurl{http://adsabs.harvard.edu/abs/1998sese.conf..729R}

\bibitem[\protect\citeauthoryear{{Rosswog}, {Liebend{\"o}rfer}, {Thielemann},
  {Davies}, {Benz} \& {Piran}}{{Rosswog} et~al.}{1999}]{rosswog99}
{Rosswog} S.,  {Liebend{\"o}rfer} M.,  {Thielemann} F.-K.,  {Davies} M.~B.,
  {Benz} W.,    {Piran} T.,  1999, A\&A, 341, 499, 
  \adsurl{http://adsabs.harvard.edu/abs/1999A\&A...341..499R}

\bibitem[\protect\citeauthoryear{{Rosswog}, {Piran} \& {Nakar}}{{Rosswog}
  et~al.}{2013}]{rosswog13a}
{Rosswog} S.,  {Piran} T.,    {Nakar} E.,  2013, MNRAS, 430, 2585,
  \adsurl{http://adsabs.harvard.edu/abs/2013MNRAS.430.2585R}

\bibitem[\protect\citeauthoryear{{Rosswog} \& {Price}}{{Rosswog} \&
  {Price}}{2007}]{rosswog07c}
{Rosswog} S.,  {Price} D.,  2007, MNRAS, 379, 915, 
  \adsurl{http://adsabs.harvard.edu/abs/2007MNRAS.379..915R}

\bibitem[\protect\citeauthoryear{{Rosswog} \& {Ramirez-Ruiz}}{{Rosswog} \&
  {Ramirez-Ruiz}}{2002}]{rosswog02b}
{Rosswog} S.,  {Ramirez-Ruiz} E.,  2002, MNRAS, 336, L7,
  \adsurl{http://adsabs.harvard.edu/abs/2002MNRAS.336L...7R}

\bibitem[\protect\citeauthoryear{{Rosswog} \& {Ramirez-Ruiz}}{{Rosswog} \&
  {Ramirez-Ruiz}}{2003}]{rosswog03b}
{Rosswog} S.,  {Ramirez-Ruiz} E.,  2003, MNRAS, 343, L36,
  \adsurl{http://adsabs.harvard.edu/abs/2003MNRAS.343L..36R}

\bibitem[\protect\citeauthoryear{{Rosswog}, {Ramirez-Ruiz} \&
  {Davies}}{{Rosswog} et~al.}{2003}]{rosswog03c}
{Rosswog} S.,  {Ramirez-Ruiz} E.,    {Davies} M.~B.,  2003, MNRAS, 345, 1077,
  \adsurl{http://adsabs.harvard.edu/abs/2003MNRAS.345.1077R}

\bibitem[\protect\citeauthoryear{{Rosswog}, {Ramirez-Ruiz}, {Hix} \&
  {Dan}}{{Rosswog} et~al.}{2008}]{rosswog08b}
{Rosswog} S.,  {Ramirez-Ruiz} E.,  {Hix} W.~R.,    {Dan} M.,  2008, Computer
  Physics Communications, 179, 184, 
  \adsurl{http://adsabs.harvard.edu/abs/2008CoPhC.179..184R}

\bibitem[\protect\citeauthoryear{{Rowlinson}, {O'Brien}, {Tanvir}, {Zhang}
  et~al.,}{{Rowlinson} et~al.}{2010}]{rowlinson10}
{Rowlinson} A., et~al., 2010, MNRAS, p.~1479,
  \adsurl{http://adsabs.harvard.edu/abs/2010MNRAS.tmp.1479R}

\bibitem[\protect\citeauthoryear{{Ruffert}, {Janka}, {Takahashi} \&
  {Schaefer}}{{Ruffert} et~al.}{1997}]{ruffert97a}
{Ruffert} M.,  {Janka} H.,  {Takahashi} K.,    {Schaefer} G.,  1997, A \& A,
  319, 122, 
  \adsurl{http://adsabs.harvard.edu/abs/1997A\&A...319..122R}

\bibitem[\protect\citeauthoryear{Schoenberg}{Schoenberg}{1946}]{schoenberg46}
Schoenberg I.~J.,  1946, Quart. Appl. Math., 4, 45

\bibitem[\protect\citeauthoryear{{Shen}, {Toki}, {Oyamatsu} \&
  {Sumiyoshi}}{{Shen} et~al.}{1998a}]{shen98a}
{Shen} H.,  {Toki} H.,  {Oyamatsu} K.,    {Sumiyoshi} K.,  1998a, Nuclear
  Physics A, 637, 435, 
  \adsurl{http://adsabs.harvard.edu/abs/1998NuPhA.637..435S}

\bibitem[\protect\citeauthoryear{{Shen}, {Toki}, {Oyamatsu} \&
  {Sumiyoshi}}{{Shen} et~al.}{1998b}]{shen98b}
{Shen} H.,  {Toki} H.,  {Oyamatsu} K.,    {Sumiyoshi} K.,  1998b, Progress of
  Theoretical Physics, 100, 1013, 
  \adsurl{http://adsabs.harvard.edu/abs/1998PThPh.100.1013S}

\bibitem[\protect\citeauthoryear{{Somiya}}{{Somiya}}{2012}]{somiya12}
{Somiya} K.,  2012, CQG, 29, 124007,
  \adsurl{http://adsabs.harvard.edu/abs/2012CQGra..29l4007S}

\bibitem[\protect\citeauthoryear{{Spergel}, {Gehrels}, {Breckinridge},
  {Donahue}, {Dressler}, {Gaudi}, {Greene}, {Guyon}, {Hirata}, {Kalirai},
  {Kasdin}, {Moos} et~al.,}{{Spergel} et~al.}{2013}]{WFIRST}
{Spergel} D. et~al. 2013, \eprint{1305.5425}

\bibitem[\protect\citeauthoryear{{Springel}}{{Springel}}{2010}]{springel10}
{Springel} V.,  2010, ARAA, 48, 391, 
  \adsurl{http://adsabs.harvard.edu/abs/2010ARA\&A..48..391S}

\bibitem[\protect\citeauthoryear{{Surman}, {McLaughlin}, {Ruffert}, {Janka} \&
  {Hix}}{{Surman} et~al.}{2008}]{surman08}
{Surman} R.,  {McLaughlin} G.~C.,  {Ruffert} M.,  {Janka} H.,    {Hix} W.~R.,
  2008, ApJL, 679, L117, 
  \adsurl{http://adsabs.harvard.edu/abs/2008ApJ...679L.117S}

\bibitem[\protect\citeauthoryear{{Tanaka} \& {Hotokezaka}}{{Tanaka} \&
  {Hotokezaka}}{2013}]{tanaka13a}
{Tanaka} M.,  {Hotokezaka} K.,  2013, ApJ, 775, 113,
  \adsurl{http://adsabs.harvard.edu/abs/2013ApJ...775..113T}

\bibitem[\protect\citeauthoryear{{Tanaka}, {Hotokezaka}, {Kyutoku}, {Wanajo},
  {Kiuchi}, {Sekiguchi} \& {Shibata}}{{Tanaka} et~al.}{2014}]{tanaka13b}
{Tanaka} M.,  {Hotokezaka} K.,  {Kyutoku} K.,  {Wanajo} S.,  {Kiuchi} K.,
  {Sekiguchi} Y.,    {Shibata} M.,  2014, ApJ, 780, 31, 
  \adsurl{http://adsabs.harvard.edu/abs/2014ApJ...780...31T}

\bibitem[\protect\citeauthoryear{{Tanvir}, {Levan}, {Fruchter}, {Hjorth},
  {Hounsell}, {Wiersema} \& {Tunnicliffe}}{{Tanvir} et~al.}{2013}]{tanvir13}
{Tanvir} N.~R.,  {Levan} A.~J.,  {Fruchter} A.~S.,  {Hjorth} J.,  {Hounsell}
  R.~A.,  {Wiersema} K.,    {Tunnicliffe} R.~L.,  2013, Nature, 500, 547,
  \adsurl{http://adsabs.harvard.edu/abs/2013Natur.500..547T}

\bibitem[\protect\citeauthoryear{{Thielemann}, {Arcones}, {K{\"a}ppeli},
  {Liebend{\"o}rfer}, {Rauscher}, {Winteler}, {Fr{\"o}hlich}, {Dillmann},
  {Fischer}, {Martinez-Pinedo}, {Langanke}, {Farouqi}, {Kratz}, {Panov} \&
  {Korneev}}{{Thielemann} et~al.}{2011}]{thielemann11}
{Thielemann} F.-K. et al.  2011, Progress in Particle and Nuclear
  Physics, 66, 346, \adsurl{http://adsabs.harvard.edu/abs/2011PrPNP..66..346T}

\bibitem[\protect\citeauthoryear{{Timmes} \& {Swesty}}{{Timmes} \&
  {Swesty}}{2000}]{timmes00a}
{Timmes} F.~X.,  {Swesty} F.~D.,  2000, ApJS, 126, 501,
  \adsurl{http://adsabs.harvard.edu/abs/2000ApJS..126..501T}

\bibitem[\protect\citeauthoryear{{Wanajo} \& {Janka}}{{Wanajo} \&
  {Janka}}{2012}]{wanajo12}
{Wanajo} S.,  {Janka} H.-T.,  2012, ApJ, 746, 180, 
  \adsurl{http://adsabs.harvard.edu/abs/2012ApJ...746..180W}

\bibitem[\protect\citeauthoryear{{Winteler}}{{Winteler}}{2012}]{winteler12}
{Winteler} C.,  2012, PhD thesis, University Basel, CH,
  \href{http://edoc.unibas.ch/29895}{edoc}

\bibitem[\protect\citeauthoryear{{Winteler}, {K{\"a}ppeli}, {Perego},
  {Arcones}, {Vasset}, {Nishimura}, {Liebend{\"o}rfer} \&
  {Thielemann}}{{Winteler} et~al.}{2012}]{winteler12b}
{Winteler} C.,  {K{\"a}ppeli} R.,  {Perego} A.,  {Arcones} A.,  {Vasset} N.,
  {Nishimura} N.,  {Liebend{\"o}rfer} M.,    {Thielemann} F.-K.,  2012, ApJL,
  750, L22, 
  \adsurl{http://adsabs.harvard.edu/abs/2012ApJ...750L..22W}

\end{thebibliography}
\hyphenation{Post-Script Sprin-ger}

\bsp

\end{document}